\documentclass[conference]{IEEEtran}

\usepackage{amsmath}
\usepackage{amssymb}
\usepackage{amsthm}
\usepackage{booktabs}
\usepackage{cite}
\usepackage{csquotes}
\usepackage[draft]{hyperref}
\usepackage{cleveref}
\usepackage[acronym,toc=false,nonumberlist]{glossaries-extra}
\usepackage{graphicx}
\usepackage{makecell}
\usepackage{pgfplots}
\usepackage{threeparttable}
\usepackage{xspace}

\usetikzlibrary{arrows.meta}
\tikzset{actor/.style={align=center, draw, rounded corners=.20cm}}
\tikzset{arc/.style={thin,-{Stealth}}}

\setabbreviationstyle[acronym]{long-short}
\newacronym{bcet}{BCET}{Best-Case Execution Time}
\newacronym{bddf}{BDDF}{Bounded Dynamic Dataflow}
\newacronym{bdf}{BDF}{Boolean-controlled Dataflow}
\newacronym{bf}{BF}{Blocking Factor}
\newacronym{biro}{BIRO}{Between Iteration Runtime Oriented}
\newacronym{biso}{BISO}{Between Iteration Statically Oriented}
\newacronym{bpdf}{BPDF}{Boolean Parametric Dataflow}
\newacronym{cddf}{CDDF}{Cyclo Dynamic Dataflow}
\newacronym{cfdf}{CFDF}{Core Functional Dataflow}
\newacronym{cf-psdf}{CF-PSDF}{Core Functional - Parameterized Synchronous Dataflow}
\newacronym{cg}{CG}{Computation Graph}
\newacronym{co}{Co}{Consistency}
\newacronym{cps}{CPS}{Cyber-Physical System}
\newacronym{cthr}{CT}{Consumption Threshold}
\newacronym{csdf}{CSDF}{Cyclo-Static Dataflow}
\newacronym{csdfa}{CSDF$^a$}{Cyclo-Static Dataflow with auto-concurrency}
\newacronym{cv-sdf}{CV-SDF}{Computer Vision-Synchronous Dataflow}
\newacronym{dec}{Dec}{Decidability}
\newacronym{del}{Del}{Delay}
\newacronym{det}{FuncDet}{Functional Determinism}
\newacronym{dfg}{DFG}{Dataflow Graph}
\newacronym{dfmocc}{DF~MoCC}{Dataflow Model of Computation and Communication}
\newacronym{dp}{DP}{Dataflow Process}
\newacronym{dpn}{DPN}{Dataflow Process Network}
\newacronym{eidf}{EIDF}{Enable-Invoke Dataflow}
\newacronym{esadf}{eSADF}{Exponentially timed Scenario-Aware Dataflow}
\newacronym{et}{ET}{Execution Time}
\newacronym{execwin}{ExecWin}{Execution Windows}
\newacronym{fifo}{FIFO}{First-In First-Out}
\newacronym{frdf}{FRDF}{Fractional Rate Dataflow}
\newacronym{fsm}{FSM}{Finite State Machine}
\newacronym{fsm-sadf}{FSM-SADF}{Finite State Machine-based Scenario-Aware Dataflow}
\newacronym{fsm-psadf}{FSM-PSADF}{Finite State Machine-based Parameterized Scenario-Aware Dataflow}
\newacronym{freq}{Freq}{Frequency}
\newacronym{gs}{GS}{Global State}
\newacronym{hcfdf}{HCFDF}{Hierarchical Core Functional Dataflow}
\newacronym{hdf}{HDF}{Heterochronous Dataflow}
\newacronym{hi}{Hi}{Hierarchy}
\newacronym{hpdf}{HPDF}{Homogeneous Parameterized Dataflow}
\newacronym{hsdf}{HSDF}{Homogeneous Synchronous Dataflow}
\newacronym{hsdfa}{HSDF$^a$}{Homogeneous Synchronous Dataflow with auto-concurrency}
\newacronym{ibsdf}{IBSDF}{Interface-Based Synchronous Dataflow}
\newacronym{idf}{IDF}{Integer-controlled Dataflow}
\newacronym{ildf}{ILDF}{Interval-rate Locally-static Dataflow}
\newacronym{inistep}{IniSteP}{Initial and Steady Phases}
\newacronym{inidisit}{IniDisIT}{Initial and Discard of Initial Tokens}
\newacronym{it}{IT}{Initial Tokens}
\newacronym{kpn}{KPN}{Kahn Process Network}
\newacronym{la}{La}{Latency}
\newacronym{li}{Li}{Liveness}
\newacronym{ma}{MA}{Markov Automata}
\newacronym{mdf}{MDF}{Multi-Dimensional FIFO}
\newacronym{mdsdf}{MDSDF}{Multi-Dimensional Synchronous Dataflow}
\newacronym{me}{Me}{Memory}
\newacronym{milp}{MILP}{Mixed Integer Linear Programming}
\newacronym{mm}{MM}{Meta-Model}
\newacronym{mmlp}{MMLP}{Min-Max Linear Programming}
\newacronym{mocc}{MoCC}{Model of Computation and Communication}
\newacronym{ooc}{OOC}{Out-of-Order Consumption}
\newacronym{pa}{Pa}{Parameters}
\newacronym{pci}{PCI}{Production and Consumption Instants}
\newacronym{pcg}{PCG}{Phased Computation Graph}
\newacronym{pcsdf}{PCSDF}{Parameterized Cyclo-Static Dataflow}
\newacronym{ph}{Ph}{Phase}
\newacronym{pimm}{PIMM}{Parameterized and Interfaced Meta-Model}
\newacronym{pisdf}{PISDF}{Parameterized and Interfaced Synchronous Dataflow}
\newacronym{ppsdf}{ppSDF}{Partially Periodic Synchronous Dataflow}
\newacronym{psdf}{PSDF}{Parameterized Synchronous Dataflow}
\newacronym{psm}{PSM}{Parameterized Set of Modes}
\newacronym{psm-cfdf}{PSM-CFDF}{Parameterized Set of Modes - Core Functional Dataflow}
\newacronym{qsc}{QSc}{Quasi-Static Schedule}
\newacronym{rai}{RaI}{Rate as Interval}
\newacronym{rdf}{RDF}{Reconfigurable Dataflow}
\newacronym{rmdf}{RMDF}{Real-time Mode-aware Dataflow}
\newacronym{rpn}{RPN}{Reactive Process Network}
\newacronym{sad}{SAD}{State-Aware Dataflow}
\newacronym{sadf}{SADF}{Scenario-Aware Dataflow}
\newacronym{stasch}{StaSch}{Statically Schedulable}
\newacronym{sco}{SCo}{Strong Consistency}
\newacronym{sdf}{SDF}{Synchronous Dataflow}
\newacronym{spbdf}{SPBDF}{Synchronous PiggyBacked Dataflow}
\newacronym{spdf}{SPDF}{Schedulable Parametric Dataflow}
\newacronym{ssdf}{SSDF}{Scalable Synchronous Dataflow}
\newacronym{swi}{SWi}{Sliding Windows}
\newacronym{tcsdf}{tCSDF}{Timed Cyclo-Static Dataflow}
\newacronym{th}{Th}{Throughput}
\newacronym{tpdf}{TPDF}{Transaction Parameterized Dataflow}
\newacronym{tsdf}{tSDF}{Timed Synchronous Dataflow}
\newacronym{uppaal}{UPPAAL}{UPPAAL}
\newacronym{vlsi}{VLSI}{Very Large Scale Integration}
\newacronym{vpdf}{VPDF}{Variable-rate Phased Dataflow}
\newacronym{vrdf}{VRDF}{Variable Rate Dataflow}
\newacronym{vsdf}{VSDF}{Synchronous Dataflow for VLSI}
\newacronym{wcet}{WCET}{Worst-Case Execution Time}
\newacronym{wiro}{WIRO}{Within Iteration Runtime Oriented}
\newacronym{wiso}{WISO}{Within Iteration Statically Oriented}
\newacronym{wsdf}{WSDF}{Windowed Synchronous Dataflow}
\newacronym{xsadf}{xSADF}{Flexible Scenario-Aware Dataflow}

\newtheorem{theorem}{Theorem}
\newtheorem{definition}[theorem]{Definition}

\newcommand{\ie}{\unskip, i.e.,\xspace}
\newcommand{\rt}{runtime\xspace}
\newcommand{\polygraph}{\textsc{PolyGraph}\xspace}
\newcommand{\eg}{\unskip, e.g.,\xspace}

\begin{document}

  \title{An Extended Survey and a Comparison Framework for Dataflow Models of Computation and Communication}

  \author{\IEEEauthorblockN{Guillaume Roumage$^\dagger$, Selma Azaiez$^\ddagger$, Cyril Faure$^\ddagger$, Stéphane Louise$^\ddagger$} \IEEEauthorblockA{$^\dagger$guillaume.roumage.research@proton.me \\ $^\ddagger$\textit{Université Paris-Saclay, CEA, List, F-91120, Palaiseau, France} \\ $^\ddagger$firstname.lastname@cea.fr}}

  \maketitle

  \begin{abstract}
    \glspl{dfmocc} is a formalism used to specify the behavior of \glspl{cps}. \glspl{dfmocc} are widely used in the design of \glspl{cps}, as they provide a high-level of abstraction to specify the system's behavior. \glspl{dfmocc} rules give semantics to a dataflow specification of a \gls{cps}, and static analysis algorithms rely on these semantics to guarantee safety properties of the dataflow specification, such as bounded memory usage and deadlock freeness. A wide range of \glspl{dfmocc} exists, each with its own characteristics and static analyses. This paper presents a survey of those \glspl{dfmocc} and a classification in eight categories. In addition, \glspl{dfmocc} are characterized by a comprehensive list of features and static analyses, which reflect their expressiveness and analyzability. Based on this characterization, a framework is proposed to compare the expressiveness and the analyzability of \glspl{dfmocc} quantitatively.
  \end{abstract}


  \section{Introduction}
  
  \glsentryfullpl{cps} are reactive systems that detect environmental shifts through sensors, process this information using computational processes, and then use the output to control actuators. CPSs range from digital signal processing systems to embedded/cloud infrastructures, soft/hard real-time systems, and even a mix of all the above. These complex systems must operate reliably without threatening their internal processes. For example, a failure of the actuator in an autonomous car can lead to catastrophic consequences such as a car crash or a pedestrian accident.
  
  Researchers and engineers seek to understand the behavior of these \glspl{cps} and ensure their safety. They develop methods to do so, such as \glspl{mocc}. A \gls{mocc} specifies rules that govern the execution of the system's specification and the communication between its components. Engineers and researchers often use sketches to conceptualize and illustrate systems and ideas in the early stages of development \cite{castrillon_dataflow_2023}. Block diagrams are a popular informal model for high-level specifications of these systems. A block diagram consists of boxes representing system components and arrows representing the relationship between components. A key advantage of this representation is that it can be fine-grained (e.g., one block specifies a processor instruction) or coarse-grained (e.g., one block is a function or even an aggregation of functions). The first mathematically grounded \gls{mocc} based on graphs was independently created by Kahn \cite{kahn_semantics_1974} and Dennis \cite{dennis_first_1974}, laying the foundations of the \gls{dfmocc} family.

  Since the seminal work of Lee and Messerschmitt \cite{lee_synchronous_1987}, the \glspl{dfmocc} family has expanded significantly, now numbering nearly 50 \gls{dfmocc} today. Each \gls{dfmocc} is a trade-off between its expressiveness -referring to the variety of systems it can specify- and its analyzability, which pertains to the static analyses it can perform. Examples of such static analyses include memory-boundedness, deadlock-freeness, feasibility and schedulability tests.

  A framework for comparing \glspl{dfmocc} has been proposed in~\cite{lee_framework_1998}. The fundamental difference between our work and theirs is that in~\cite{lee_framework_1998}, a \emph{denotational semantics} is provided. Actors, channels, and tokens are described in terms of mathematics objects. Our approach has \emph{operational semantics} \ie we described the computation capabilities rather than mapping them to a mathematical object. The work of~\cite{bouakaz_survey_2017} is quite similar but more focused on parametric \glspl{dfmocc}.

  \subsubsection{Contributions}

  The contribution of this paper is an extension of the \glspl{dfmocc} survey of \cite{roumage_survey_2022}. The classification categories have been refined, and the features and static analyses have been updated. As \glspl{dfmocc} are a trade-off between their expressiveness and their analyzability, this paper proposes a methodology based on system designer needs to compute the expressiveness and the analyzability score. These scores are two numerical values to evaluate the expressiveness and the analyzability of a \gls{dfmocc}.

  \subsubsection{Paper organization}

  This paper starts by presenting similarities between all \glspl{dfmocc} in \cref{sec:dataflow-principles}. In \cref{sec:features} and \cref{sec:analyzability}, the features and static analyses of our framework are presented, respectively. Some insights about the Turing completeness of \gls{dfmocc} are given in \cref{sec:turing-completeness}, and \cref{sec:classification} presents our classification and its eight categories. A protocol to compute the expressiveness and analyzability score of a \gls{dfmocc} is presented in \cref{sec:hierarchy}. Finally, \cref{sec:conclusion} concludes the paper.

  \section{Dataflow Principles}

  \label{sec:dataflow-principles}

  Although each \gls{dfmocc} defines its own semantics for specifying and analyzing \glspl{cps}, they share the same background. They are mainly based on the concept of \gls{dfg}.

  \begin{definition}[Dataflow Graph]
    A \gls{dfg} is a directed graph $G = (V, E)$ that specifies a system where vertices represent actors of the system, and edges represent communication channels between actors. A \gls{dfg} specifies a system, and the semantics of this \gls{dfg} is given by the \gls{dfmocc} used. We also say that a \gls{dfg} is an \emph{instance} or a \emph{specification} of a \gls{dfmocc}.
  \end{definition}

  An \emph{actor} is a computational unit that both produces and consumes data every time it is executed \ie every time it executes a \emph{job}. The atomic amount of data exchanged is known as a \emph{token}. Usually, tokens within channels are produced and consumed with a \gls{fifo} policy, meaning they are consumed in the order in which they are produced. The internal behavior of an actor may only be partially known by systems designers, and sometimes, it is not at all. Hence, actors lie between \enquote{white boxes} and \enquote{black boxes}. The minimum amount of information system designers need is only the number of tokens produced and consumed by an actor each time it is executed.

  \begin{definition}[Channel]
    A channel is formally defined as a tuple $c_i = (v_j, v_k, n_{ij}, n_{ik}, [c_i])$ where $v_j$ is the \emph{producer} actor, $v_k$ is the \emph{consumer} actor, $n_{ij}$ is the \emph{production rate} \ie the number of tokens produced by $v_j$ on $c_i$ each time it executes a job, $n_{ik}$ is the \emph{consumption rate} \ie the number of tokens consumed by $v_k$ on $c_i$ each time it executes a job, and $[c_i]$ is the \emph{initial tokens} of $c_i$.
  \end{definition}

  \glspl{dfmocc} can be classified into \emph{functional deterministic} and \emph{non-functional deterministic} ones. A \gls{dfmocc} has a functional determinism if the output of the actors of the \gls{dfg} solely depends on their inputs, independently from external factors such as time or randomness. The \emph{temporal determinism} of a \gls{dfmocc} asserts that the execution windows of the actors are fixed and do not vary from one execution to another. In this paper, we will focus on functional determinism; in this paper, a deterministic \gls{dfmocc} is a \gls{dfmocc} with a functional determinism.
  
  Although both functional deterministic and non-functional deterministic \glspl{dfmocc} have static analysis algorithms, the former ones are more accurate. A static analysis permits the prediction of some facets of the \rt behavior of a \gls{cps} at compile-time. Those static analyses often rely on the topology of the CPS's specification and the communication pattern among actors. Both of these information are stored in the \emph{topology matrix}.

  \begin{definition}[Topology matrix]
    The \emph{topology matrix} of a \gls{dfg} $G = (V, E)$ is denoted $G_\Gamma = (\gamma_{ij})$. The range rate (e.g, $\mathbb{N}^*, \mathbb{N}, \mathbb{Q}^*$) of $\gamma_{ij}$ depends on which \gls{dfmocc} defines the semantics of the \gls{dfg}. In any case, $\gamma_{ij}$ is the consumption/production rate of the actor $v_j$ from/to the channel $c_i$. If $v_j$ consumes/produces, $\gamma_{ij}$ is negative/positive.
  \end{definition}

  \begin{definition}[Actor]
    An actor is formally defined as a tuple $v_i = (I_i, O_i)$ where $I_i \subseteq E$ is the set of input channels of $v_i$ and $O_i \subseteq E$ is the set of output channels of $v_i$. Note that $I_i$ and $O_i$ may be empty. If $I_i$ is empty, $v_i$ is a \emph{source} actor or a \emph{sensor}, and if $O_i$ is empty, $v_i$ is a \emph{sink} actor or an \emph{actuator}.
  \end{definition}

  Unless stated otherwise, an actor can execute if and only if the number of tokens in its input channels is greater than the consumption rate of the channel. Whenever an actor executes, it writes/reads an amount of token equal to the production/consumption rate of the channel. Write operations are non-blocking \ie an actor can produce tokens whenever it executes, and read operations are blocking \ie an actor consumes tokens only if it executes and if there are enough tokens in the input channels.

  \begin{definition}[Iteration]
    An iteration of a DFG is a partially ordered collection of actors' executions that keeps the token distribution of the DFG unchanged.
  \end{definition}

  \section{Characterization of the expressiveness of DF MoCCs}

  \label{sec:features}

  Each \gls{dfmocc} can specify different characteristics of CPSs, which is closely tied to the attainable \emph{model fidelity} of a \gls{dfmocc}. Depending on the CPS being specified, not all \glspl{dfmocc} are suitable. We defined those characteristics as \emph{features}, which we will detail in the following sections.
  
  We conducted a survey to identify the most important features. As a result, some features are shared among multiple \glspl{dfmocc}, while others are unique to a specific \gls{dfmocc}. 
  
  Features are classified into three categories based on how they manifest in the \gls{dfmocc}. The first category (cf. \cref{sec:present-absent-features}) presents 18 features that are either \emph{present-or-absent} \eg the possibility of specifying frequency constraint. The second category (cf. \cref{sec:range-rate}) involves evaluating the range of the production and consumption rates, and the third category (cf. \cref{sec:rate-topology-updates}) involves assessing the rate and topology updates of the \gls{dfmocc}. Last but not least, features are not independent; those dependencies are detailed in \cref{sec:dependencies-features}.

  \subsection{Comprehensive List of Present-or-Absent Features}

  \label{sec:present-absent-features}

  There are 18 \emph{present-or-absent} features for \glspl{dfmocc}. The following sections will detail those features in a lexicographic order and briefly explain each feature.

  \subsubsection{\underline{B}locking \underline{F}actor (BF)}

  Actors can consume and produce any multiple of their consumption and production rates. That multiple is called the \emph{blocking factor}.

  \subsubsection{\underline{C}onsumption \underline{T}hreshold (CT)}

  The number of tokens in a channel must exceed a threshold for an actor to consume them. The consumption threshold is usually different from the consumption rate.

  \subsubsection{\underline{G}lobal \underline{S}tate (GS)}

  A key-value structure is shared among actors.

  \subsubsection{\underline{Hi}erarchy (\glsentryshort{hi})}

  Compositionality can be achieved by associating a subgraph to an actor.

  \subsubsection{\underline{Ini}tial and \underline{Ste}ady \underline{P}hases (IniSteP)}

  Actors can have initial phases followed by cyclic ones (cf. the \glsentrylong{ph} feature below).

  \subsubsection{\underline{I}nitial \underline{T}okens (IT)}

  Tokens can be stored in channels' buffer before the start of the execution of the system.

  \subsubsection{\underline{Ini}tialization and \underline{Dis}card of \underline{I}nitial \underline{T}okens (IniDisIT)}

  An explicit mechanism initializes and discards initial tokens at the start and the end of each iteration of the \gls{dfg}.

  \subsubsection{\underline{M}ulti-\underline{D}imensional \underline{F}IFO (MDF)}

  Instead of being a single queue, channels' buffer can be described as multi-dimensional lattices.

  \subsubsection{\underline{M}eta-\underline{M}odel (MM)}

  Enhance the expressiveness of a non-meta-model \gls{dfmocc} with additional rules.

  \subsubsection{\underline{O}ut-of-\underline{O}rder \underline{C}onsumption (OOC)}

  Out-of-order consumption of tokens is allowed \ie the FIFO policy is not enforced.

  \subsubsection{\underline{Pa}rameters (Pa)}

  Production and consumption rates are not necessarily fixed scalars.

  \subsubsection{\underline{Ph}ases (Ph)}

  Production and consumption rates vary from one job to another.

  \subsubsection{\underline{R}ate \underline{a}s \underline{I}nterval (RaI)}

  Production and consumption rates are given as intervals.

  \subsubsection{\underline{E}xecution \underline{T}ime (ET)}

  Actors can have a non-null execution time.

  \subsubsection{\underline{Freq}uency (Freq)}

  Actors can have a frequency constraint.

  \subsubsection{\underline{Del}ay (Del)}

  Actors can have a delay constraint \ie the time of their first execution is postponed for a given amount of time.

  \subsubsection{\underline{P}roduction and \underline{C}onsumption \underline{I}nstants (PCI)}

  Specify the time instants when tokens can be produced and consumed.

  \subsubsection{\underline{S}liding \underline{W}indow (SW)}

  Tokens can be consumed with sliding windows through the channels' buffer.

  \subsection{Range Rate}

  \label{sec:range-rate}

  The definition of the topology matrix states that the production and consumption range rates depend on the underlying \gls{dfmocc}. The \emph{range rate} is the set of values the production and consumption rates can take. It ranges from the simplest, which is the singleton $\{1\}$, to more complex domains such as $\mathbb{Q}^*$. The former is less expressive as it implies that the production and consumption rates are fixed at $1$. The latter is more expressive as it allows the specification of any positive rational number \ie an actor produces and consumes a \emph{fractional number} of tokens. The range rate is an element within the following set: $\{ \{ 1 \}, \mathbb{N}^*, \mathbb{N}, \mathbb{Q}^*, \Omega \}$. The semantics of the $\Omega$ will be detailed later. Intuitively, it means that \enquote{any type of object can be produced and consumed}.

  \subsection{Rate and Topology Updates}

  \label{sec:rate-topology-updates}

  The rate and topology updates are measured in two dimensions: the instants at which they can change (the \emph{when}) and the type of the procedure to define new rates (the \emph{how}). Rate updates may induce topology updates. For instance, setting a rate to zero means that no tokens are produced or consumed. Therefore, the associated channel is no longer used if that rate is zero, and the topology is updated.

  First, the \emph{when}. Rate and topology updates can change \emph{between} iterations or \emph{within} iterations of the \gls{dfg}. When it occurs between iterations, it implies a static behavior of the \gls{dfg} during an iteration, but it is more restrictive and less expressive as an update within iterations. Second, the \emph{how}. New rate values can be determined either at \emph{compile-time} or at \emph  {\rt}. If the set of possible values for rates is fixed at compile-time, the \gls{dfg} is said to be \emph{statically-oriented}. If this set varies at \rt, the \gls{dfg} is \emph{\rt-oriented}.

  These two dimensions can be combined: the \emph{when} is either \emph{between} or \emph{within} iterations, and the \emph{how} is either \emph{statically-oriented} or \emph{\rt-oriented}. Thus, the rate and topology updates can be classified into four categories: \gls{biro}, \gls{biso}, \gls{wiro}, and \gls{wiso}. A combination of these behaviors is possible if, for instance, some range rates are fixed at compile-time, and others are determined at \rt.

  \subsection{Dependencies between Features}

  \label{sec:dependencies-features}

  The expressiveness of \glspl{dfmocc} has been classified into the \emph{range rate} feature, the \emph{rate and topology updates} feature, and 18 \emph{present-or-absent} features, namely \glsentryfull{bf}, \glsentryfull{cthr}, \glsentryfull{gs}, \glsentryfull{hi}, \glsentryfull{inistep}, \glsentryfull{it}, \glsentryfull{inidisit}, \glsentryfull{mdf}, \glsentryfull{mm}, \glsentryfull{ooc}, \glsentryfull{pa}, \glsentryfull{ph}, \glsentryfull{rai}, \glsentryfull{et}, \glsentryfull{freq}, \glsentryfull{del}, \glsentryfull{pci}, \glsentryfull{swi}.
  
  Those features are not independent. Those dependencies are presented under the form \enquote{\emph{Feature $i_1$} + \dots + \emph{Feature $i_n$} $\to$ \emph{Feature $j_1$} + \dots + {Feature $j_n$}} such that if there is the set of features of the left-hand side, then the set of features of the right-hand side is also present.

  \begin{itemize}
    \item \glsentryshort{inidisit} $\to$ \glsentryshort{it}: The presence of a mechanism to explicitly handle the initialization and discard of initial tokens necessarily implies the presence of initial tokens.
    \item $range~rate = \mathbb{Q}^* \to$ \glsentryshort{ph}: The semantics of a rational rate $p/q$ is that an actor produces and consumes $p$ tokens over $q$ consecutive executions, so the number of tokens produced and consumed may vary from one execution to another.
  \end{itemize}

  \section{Characterization of the analyzability of DF MoCCs}

  \label{sec:analyzability}

  \subsection{Comprehensive List of Static Analyses}

  There are 11 static analyses that can be performed on a \gls{dfg}. The following sections will detail those static analyses in a lexicographic order and provide a brief explanation of each static analysis.

  \subsubsection{\underline{Co}nsistency (\glsentryshort{co})}

  The existence of an execution in bounded memory is decidable at compile-time.

  \subsubsection{\underline{Dec}idability (\glsentryshort{dec})}

  The feasibility of an execution is provable at compile-time.

  \subsubsection{\underline{Func}tional \underline{Det}erminism (\glsentryshort{det})}

  The same sequence of outputs is produced in response to a given sequence of inputs. In other words, outputs do not depend on external factors such as time or randomness.

  \subsubsection{\underline{Exec}ution \underline{Win}dows (\glsentryshort{execwin})}

  The execution window of the actors can be computed at compile-time.

  \subsubsection{\underline{La}tency (\glsentryshort{la})}

  The latency of an execution can be computed at compile-time.

  \subsubsection{\underline{Li}veness (\glsentryshort{li})}

  The existence of a deadlock-free execution is decidable at compile-time.

  \subsubsection{\underline{Me}mory (\glsentryshort{me})}
  
  The memory footprint of an execution can be computed at compile-time.

  \subsubsection{\underline{Q}uasi-staticically \underline{Sche}dulable (\glsentryshort{qsc})}

  A quasi-static schedule can be derived at compile-time.

  \subsubsection{\underline{Sta}tically \underline{Sch}edulable (\glsentryshort{stasch})}

  A static schedule can be derived at compile-time. Usually, there is no limit on the number of cores executing the system. While common in the context of \glspl{dfmocc}, this assumption is not so often found in other communities.

  \subsubsection{\underline{S}trong \underline{Co}nsistency (\glsentryshort{sco})}

  The boundedness of all executions is decidable at compile-time.

  \subsubsection{\underline{Th}roughput (\glsentryshort{th})}

  The throughput of an execution can be computed at compile-time.

  \subsection{Dependencies between Static Analyses}

  The analyzability of \glspl{dfmocc} has been classified into 11 properties: \glsentryfull{co}, \glsentryfull{dec}, \glsentryfull{det}, \glsentryfull{execwin}, \glsentryfull{la}, \glsentryfull{li}, \glsentryfull{me}, \glsentryfull{qsc}, \glsentryfull{stasch}, \glsentryfull{sco}, and \glsentryfull{th}. 
  
  Those properties are not independent. Those dependencies are presented under the form \enquote{\emph{Prop $i_1$} + \dots + \emph{Prop $i_n$} $\to$ \emph{Prop $j_1$} + \dots + {Prop $j_n$}} such that if there is the set of properties of the left-hand side, then the set of properties of the right-hand side is also present.

  \begin{itemize}
    \item \glsentryshort{stasch} $\to$ \glsentryshort{dec} + \glsentryshort{co} + \glsentryshort{li} and \glsentryshort{qsc} $\to$ \glsentryshort{dec} + \glsentryshort{co} + \glsentryshort{li}: The derivation of a static/quasi-static schedule at compile-time implies that the feasibility is provable, so there exists at least one deadlock-free execution in bounded memory.
    \item \glsentryshort{dec} $\to$ \glsentryshort{co} + \glsentryshort{li}: The feasibility of a system implies that at least one deadlock-free execution exists in bounded memory.
    \item \glsentryshort{sco} $\to$ \glsentryshort{co}: The guarantee of memory boundedness of all executions necessarily implies the existence of at least one memory-bounded execution.
    \item \glsentryshort{li} $\to$ \glsentryshort{co} + \glsentryshort{dec}: The liveness of a dataflow specification implies that there exist a deadlock-free execution, so it implies also the consistency and the feasibility of the specification.
    \item \glsentryshort{th} $\to$ \glsentryshort{et} and \glsentryshort{la} $\to$ \glsentryshort{et}: Throughput and latency evaluate quantitatively some performance of the system over time, so those two properties imply that actors have execution times.
    \item \glsentryshort{execwin} $\to$ \glsentryshort{et}: The computation of execution window length requires the execution time of actors.
    \item \glsentryshort{me} $\to$ \glsentryshort{stasch} and \glsentryshort{me} $\to$ \glsentryshort{qsc}: A memory footprint computation needs to know how the channel buffers are used, which implies knowing a static schedule or a quasi-static schedule.
  \end{itemize}

  \section{Characterization of the Turing-Completeness}

  \label{sec:turing-completeness}

  Some \glspl{dfmocc} are shown to be Turing complete \cite{hopcroft_introduction_2001} \ie they can simulate a Turing machine. In particular, the \emph{halting problem} is undecidable for Turing machines. Thus, it is impossible to determine if an execution of an instance of a Turing-complete \gls{dfmocc} will stop or not. Consequently, static analyses such as consistency or liveness are generally undecidable for Turing-complete \glspl{dfmocc}. However, some Turing-complete \glspl{dfmocc} propose restrictions and conditions on their instances to determine safety properties. These will be detailed for \glspl{dfmocc} involved.

  \section{The Classification}

  \label{sec:classification}

  We propose a new classification of \glspl{dfmocc} that is different from the traditional \emph{static}, \emph{reconfigurable}, and \emph{dynamic} \glspl{dfmocc}. We propose to classify \glspl{dfmocc} for \glspl{cps} design and verification into eight categories as follows: \nameref{sec:sdf-and-related-dfmoccs} (\cref{tab:sdf-and-related-dfmoccs}), \nameref{sec:phased-based-dfmoccs} (\cref{tab:phased-based-dfmoccs}), \nameref{sec:timed-based-dfmoccs} (\cref{tab:timed-based-dfmoccs}), \nameref{sec:boolean-based-dfmoccs} (\cref{tab:boolean-based-dfmoccs}), \nameref{sec:scenario-based-dfmoccs} (\cref{tab:scenario-based-dfmoccs}), \nameref{sec:meta-model-dfmoccs} (\cref{tab:meta-model-dfmoccs}), \nameref{sec:dfmocc-with-enable-and-invoke-capabilities} (\cref{tab:dfmocc-with-enable-and-invoke-capabilities}) and \nameref{sec:process-network-based-dfmocc} (\cref{tab:process-network-based-dfmocc}).

  \subsection{Synchronous Dataflow and Related DF MoCCs}

  \label{sec:sdf-and-related-dfmoccs}

  This category includes \glspl{dfmocc} which are very similar to the \gls{sdf}. Those similar \glspl{dfmocc} can differ from \gls{sdf} by the range rate, the rate and topology updates, or even have parametric rates. For each \gls{dfmocc} of this category, we will detail the differences with \gls{sdf}.

  \subsubsection{SDF}

  The roots of \gls{dfmocc} can be traced back to their earliest form in~\cite{dennis_first_1974}. However, the analyzability of this model was limited. \gls{sdf}~\cite{lee_synchronous_1987} has paved the way for the current dataflow paradigm. An \gls{sdf} specification is a \gls{dfg} where the production and consumption rates of the channels belong to $\mathbb{N}^*$. \gls{sdf} has consistency and liveness checking algorithms, and a static schedule can be derived~\cite{lee_static_1987} while optimizing the memory footprint~\cite{fradet_sequential_2023}. Reference~\cite{ghamarian_liveness_2006} extends the usual consistency property to the strong consistency property \ie deciding whether \emph{all} executions are bounded. \gls{sdf} has been extensively researched for its ability to specify various applications, and many works have delved into the memory footprint minimization problem. For instance, a shared buffer memory model is examined in~\cite{murthy_shared_2000}, and a buffer merging technique is proposed in~\cite{desnos_buffer_2015}. While the memory footprint minimization problem is NP-complete~\cite{bhattacharyya_software_1996}, an exact method to solve this problem is proposed in~\cite{geilen_minimising_2005} with model-checking. Another approach presented in~\cite{ade_data_1997} involves arithmetic manipulations between production and consumption rates and the number of initial tokens to provide the minimum required buffer size, yielding a deadlock-free execution.

  \subsubsection{HSDF}

  The rates' values of an \gls{sdf} specification can be restricted to $\{1\}$ to yield an \gls{hsdf} specification~\cite{lee_synchronous_1987}. The static analysis algorithms of \gls{sdf} also apply to \gls{hsdf}.

  \subsubsection{SSDF}

  \gls{ssdf}~\cite{ritz_high_1992} aims to improve the implementation efficiency of \gls{sdf} specifications. It introduces the concept of the \emph{blocking factor}, allowing actors to produce and consume a positive multiple of their consumption and production rates at each execution. This reduces context-switch overhead by performing multiple computations at once. The value of blocking factors is defined at compile-time and cannot be modified at \rt. A detailed investigation into scheduling strategies to maximize throughput and optimize memory footprint has been conducted~\cite{ritz_optimum_1993,ritz_scheduling_1995}.

  \subsubsection{BDDF}

  \gls{bddf}~\cite{pankert_dynamic_1994} extends \gls{ssdf} by introducing dynamic and upper-bounded ports for a set of actors, allowing their rates to change at \rt up to a maximum value. This addresses some limitations of \gls{ssdf}, such as the possibility of adjusting consumption or production rates to zero for an uncertain number of executions. The value of dynamic ports depends only on current and previous values token values. Therefore, \gls{bddf} is deterministic. \gls{bddf} includes topology updates, with a \gls{fsm} used to specify topology configuration, where each state defines a set of connected actors. Note that actors with an \gls{ssdf} semantics can coexist within a \gls{bddf} specification. In addition, given that dynamic ports are upper-bounded, a consistency analysis similar to \gls{sdf}~\cite{lee_static_1987} can be performed.

  \subsubsection{CG}

  \glspl{cg}~\cite{karp_properties_1966} are more general than \gls{sdf}. In a \gls{cg}, each channel is also associated with a consumption threshold. Thus, an actor can execute if the number of tokens in its input channel exceeds both the consumption rate \emph{and} the threshold. The authors of~\cite{karp_properties_1966} studied safety properties different than the usual ones, including the \emph{determinacy} \ie the fact that any execution leads to the same result, and the \emph{termination} \ie the fact that any execution terminates. This latter is quite different from the liveness, which asserts that a system \emph{can} actually have a non-terminating execution. As the termination property is outside the scope of our interest, we do not consider it. Reference~\cite[Section 4.7]{bhattacharyya_software_1996} provides a condition to generate a static schedule of a \gls{cg}. The main idea is to find a schedule of a semantically equivalent \gls{sdf} specification.

  \subsubsection{SPDF}

  \gls{spdf}~\cite{fradet_spdf_2012} is an extended version of \gls{sdf} that incorporates a set of parameters within the range of $\mathbb{N}^*$ and defined at compile-time. These parameters are communicated through a dedicated network integrated at the top of an \gls{spdf} specification and are subject to change within an iteration. \gls{spdf} is associated with static analysis algorithms to check consistency and liveness, and a quasi-static schedule can be derived.

  \subsubsection{MDSDF}

  \gls{mdsdf}~\cite{murthy_multidimensional_2002} extends \gls{sdf} by defining the number of tokens produced and consumed as multi-dimensional lattices. \gls{mdsdf} is suitable for modeling signal processing applications, such as image processing. Although tokens are multi-dimensional, we consider that the range rate is $\mathbb{N}^*$ as each dimension is given as a positive integer. A schedule of an \gls{mdsdf} specification can be determined at compile-time~\cite{murthy_scheduling_1996}. Reference~\cite{lee_multidimensional_1993} provides the initial version of \gls{mdsdf}, where tokens have a rectangular shape, and~\cite{murthy_multidimensional_2002} goes one step further by considering an arbitrary lattice shape. The token shape is essential as it impacts how actors should read and write them.

  \subsubsection{WSDF}

  \gls{wsdf}~\cite{keinert_windowed_2011} extends \gls{mdsdf} by allowing tokens to be consumed with sliding windows. A token is consumed with a specific sampling pattern through a predefined window. A schedule of a \gls{wsdf} specification can be derived at compile-time.

  \subsubsection{IBSDF}

  \gls{ibsdf}~\cite{piat_interfacebased_2009} is a hierarchical extension of \gls{sdf}. An \gls{ibsdf} specification is an \gls{sdf} specification with a source and sink node surrounding it. They both behave as an interface to the environment, which eases the hierarchical construction of an \gls{ibsdf} specification. Each level of the hierarchy is statically analyzable.

  \subsubsection{RDF}

  \gls{rdf}~\cite{fradet_rdf_2022} extends \gls{sdf} by incorporating a \emph{controller} that dictates how and when an \gls{sdf} specification may be reconfigured. Graph rewrite rules are enforced when particular \rt criteria are met, such as throughput/buffer occupancy dropping/above a given threshold. \gls{rdf} ensures the consistency and liveness of the original \gls{sdf} specification and all potential transformations. All possible reconfigurations of an \gls{rdf} specification do not have to be explicitly stated at design time, and their number can be arbitrarily large or even unbounded.

  \subsubsection{CV-SDF}

  \gls{cv-sdf}~\cite{stichling_cvsdf_2002} extends \gls{sdf} to simplify the modeling and analysis of computer vision applications. According to the authors, computer vision systems usually have specific requirements when they execute, such as processing frames into chunks, accessing the neighborhood of a pixel, or accessing the previous frame. They propose a new buffer structure (which no longer acts as a \gls{fifo}) along with a special consumption rate format that allows the modeling of those requirements. A schedule, as well as the memory usage, can be computed at compile-time.

  \subsubsection{SPBDF}

  \gls{spbdf}~\cite{park_extended_2002} extends \gls{sdf} by providing a global state. Two types of actors are introduced to handle the update values inside the global state; the first creates the update request, and the second updates the global state. The second actor is the only one who can write in the global state. Other actors are allowed to read the global state when they execute. The authors of~\cite{park_extended_2002} provide necessary and sufficient conditions for the consistency of the global state by proving that the global state is written only once per iteration. A static schedule is derivable based on the scheduling techniques of \gls{sdf}, and the memory footprint of the global state is computable.

  \subsubsection{HDF}

  \gls{hdf}~\cite{girault_hierarchical_1999} studied the combination between \glspl{fsm} and \glspl{dfg}. This approach allows the refinement of an actor of a \gls{dfg} into an \gls{fsm} and conversely. The questions of consistency and liveness are decidable, but it is usually impractical to compute static schedules.

  \begin{table*}[htbp]
    \centering
    \caption{Features and static analyses of \nameref{sec:sdf-and-related-dfmoccs}.}
    \label{tab:sdf-and-related-dfmoccs}
    \begin{threeparttable}[htbp]
      \begin{tabular}{|c|c|c|c|c|c|c|}
        \toprule
        \textbf{\glsentryshort{dfmocc}} & \makecell{\textbf{Rate} \\ \textbf{updates}} & \makecell{\textbf{Topology} \\ \textbf{updates}} & \makecell{\textbf{Range} \\ \textbf{rate}} & \textbf{Features} & \textbf{Static analyses} & \makecell{\textbf{Turing} \\ \textbf{complete}} \\
        \midrule
        \glsentryshort{sdf}~\cite{lee_synchronous_1987} & never & never & $\mathbb{N}^*$ & \glsentryshort{it} & \makecell{\glsentryshort{co}~\cite{lee_synchronous_1987,lee_static_1987,ghamarian_liveness_2006,fradet_sequential_2023}; \\ \glsentryshort{dec}~\cite{lee_static_1987,fradet_sequential_2023}; \glsentryshort{det}~\cite{lee_synchronous_1987}; \\ \glsentryshort{li}~\cite{lee_synchronous_1987,lee_static_1987,ade_data_1997,ghamarian_liveness_2006,fradet_sequential_2023}; \\ \glsentryshort{me}~\cite{ade_data_1997,murthy_shared_2000,geilen_minimising_2005,desnos_buffer_2015,fradet_sequential_2023}; \\ \glsentryshort{sco}~\cite{ghamarian_liveness_2006}; \\ \glsentryshort{stasch}~\cite{lee_static_1987,fradet_sequential_2023}} & $\circ$ \\
        \midrule
        \glsentryshort{hsdf}~\cite{lee_synchronous_1987} & never & never & $\{1\}$ & \glsentryshort{it} & \makecell{\glsentryshort{co}~\cite{lee_synchronous_1987,lee_static_1987,ghamarian_liveness_2006,fradet_sequential_2023}; \\ \glsentryshort{dec}~\cite{lee_static_1987,fradet_sequential_2023}; \glsentryshort{det}~\cite{lee_synchronous_1987}; \\ \glsentryshort{li}~\cite{lee_synchronous_1987,lee_static_1987,ade_data_1997,ghamarian_liveness_2006,fradet_sequential_2023}; \\ \glsentryshort{me}~\cite{ade_data_1997,murthy_shared_2000,geilen_minimising_2005,desnos_buffer_2015,fradet_sequential_2023}; \\ \glsentryshort{sco}~\cite{ghamarian_liveness_2006}; \\ \glsentryshort{stasch}~\cite{lee_static_1987,fradet_sequential_2023}} & $\circ$ \\
        \midrule
        \glsentryshort{ssdf}~\cite{ritz_high_1992} & never & never & $\mathbb{N}^*$ & \makecell{\glsentryshort{bf} \\ \glsentryshort{it}} & \makecell{\glsentryshort{co}~\cite{ritz_high_1992,ritz_optimum_1993}; \glsentryshort{dec}~\cite{ritz_high_1992,ritz_optimum_1993}; \\ \glsentryshort{det}~\cite{ritz_high_1992}; \glsentryshort{li}~\cite{ritz_high_1992,ritz_optimum_1993}; \\ \glsentryshort{me}~\cite{ritz_high_1992,ritz_optimum_1993,ritz_scheduling_1995}; \\ \glsentryshort{stasch}~\cite{ritz_high_1992,ritz_optimum_1993,ritz_scheduling_1995}} & $\circ$ \\
        \midrule
        \glsentryshort{bddf}~\cite{pankert_dynamic_1994} &  \makecell{never\tnote{1} + \\ \glsentryshort{wiro}} & \makecell{never\tnote{1} + \\ \glsentryshort{wiro}} & $\mathbb{N}$ & \makecell{\glsentryshort{bf}\tnote{1} \\ \glsentryshort{it}} & \makecell{\glsentryshort{co}~\cite{pankert_dynamic_1994}; \glsentryshort{dec}\tnote{1}~\cite{ritz_high_1992,ritz_optimum_1993}; \\ \glsentryshort{det}~\cite{pankert_dynamic_1994}; \glsentryshort{me}\tnote{1}~\cite{ritz_high_1992,ritz_optimum_1993,ritz_scheduling_1995}; \\ \glsentryshort{stasch}\tnote{1}~\cite{ritz_high_1992,ritz_optimum_1993}} & $\circ$ \\
        \midrule
        \glsentryshort{cg}~\cite{karp_properties_1966} & never & never & $\mathbb{N}^*$ & \makecell{\glsentryshort{cthr} \\ \glsentryshort{it} }& \makecell{\glsentryshort{co}~\cite{bhattacharyya_software_1996}; \glsentryshort{dec}~\cite{bhattacharyya_software_1996}; \\ \glsentryshort{det}~\cite{karp_properties_1966}; \glsentryshort{li}~\cite{bhattacharyya_software_1996}; \\ \glsentryshort{stasch}~\cite{bhattacharyya_software_1996}} & $\circ$ \\
        \midrule
        \glsentryshort{spdf}~\cite{fradet_spdf_2012} & \glsentryshort{wiso} & never & $\mathbb{N}^*$ & \makecell{\glsentryshort{it} \\ \glsentryshort{pa}} & \makecell{\glsentryshort{co}~\cite{fradet_spdf_2012}; \glsentryshort{dec}~\cite{fradet_spdf_2012}; \\ \glsentryshort{det}~\cite{fradet_spdf_2012}; \glsentryshort{li}~\cite{fradet_spdf_2012}; \\ \glsentryshort{qsc}~\cite{fradet_spdf_2012}} & $\circ$ \\
        \midrule
        \glsentryshort{mdsdf}~\cite{murthy_multidimensional_2002} & never & never & $\mathbb{N}^*$& \makecell{\glsentryshort{it} \\ \glsentryshort{mdf}} & \makecell{\glsentryshort{co}~\cite{murthy_scheduling_1996,murthy_multidimensional_2002}; \glsentryshort{dec}~\cite{murthy_scheduling_1996,murthy_multidimensional_2002}; \\ \glsentryshort{det}~\cite{murthy_scheduling_1996,murthy_multidimensional_2002}; \glsentryshort{li}~\cite{murthy_scheduling_1996,murthy_multidimensional_2002}; \\ \glsentryshort{stasch}~\cite{murthy_scheduling_1996,murthy_multidimensional_2002}} & $\circ$ \\
        \midrule
        \glsentryshort{wsdf}~\cite{keinert_windowed_2011} & never & never & $\mathbb{N}^*$ & \makecell {\glsentryshort{it} \\ \glsentryshort{mdf} \\ \glsentryshort{swi}} & \makecell{\glsentryshort{co}~\cite{keinert_windowed_2011}; \glsentryshort{dec}~\cite{keinert_windowed_2011}; \\ \glsentryshort{det}~\cite{keinert_windowed_2011}; \glsentryshort{li}~\cite{keinert_windowed_2011}; \\ \glsentryshort{stasch}~\cite{keinert_windowed_2011}} & $\circ$ \\
        \midrule
        \glsentryshort{ibsdf}~\cite{piat_interfacebased_2009} & never & never & $\mathbb{N}^*$ & \makecell{\glsentryshort{hi} \\ \glsentryshort{it}} & \makecell{\glsentryshort{co}~\cite{piat_interfacebased_2009}; \glsentryshort{dec}~\cite{piat_interfacebased_2009}; \\ \glsentryshort{det}~\cite{piat_interfacebased_2009}; \glsentryshort{li}~\cite{piat_interfacebased_2009}; \\ \glsentryshort{stasch}~\cite{piat_interfacebased_2009}} & $\circ$ \\
        \midrule
        \glsentryshort{rdf}~\cite{fradet_rdf_2022} & \glsentryshort{biro} & \glsentryshort{biro} & $\mathbb{N}^*$ & \glsentryshort{it} & \makecell{\glsentryshort{co}~\cite{fradet_rdf_2022}; \glsentryshort{dec}~\cite{fradet_rdf_2022}; \\ \glsentryshort{det}~\cite{fradet_rdf_2022}; \glsentryshort{li}~\cite{fradet_rdf_2022}} & $\circ$ \\
        \midrule
        \glsentryshort{cv-sdf}~\cite{stichling_cvsdf_2002} & never & never & $\mathbb{N}^*$ & \glsentryshort{it} & \makecell{\glsentryshort{co}~\cite{stichling_cvsdf_2002}; \glsentryshort{dec}~\cite{stichling_cvsdf_2002}; \\ \glsentryshort{det}~\cite{stichling_cvsdf_2002}; \glsentryshort{li}~\cite{stichling_cvsdf_2002}; \\ \glsentryshort{me}~\cite{stichling_cvsdf_2002}; \glsentryshort{stasch}~\cite{stichling_cvsdf_2002}} & $\circ$ \\
        \midrule
        \glsentryshort{spbdf}~\cite{park_extended_2002} & never & never & $\mathbb{N}^*$ & \makecell{\glsentryshort{gs} \\ \glsentryshort{it}} & \makecell{\glsentryshort{co}~\cite{park_extended_2002}; \glsentryshort{dec}~\cite{park_extended_2002}; \\ \glsentryshort{det}~\cite{park_extended_2002};  \glsentryshort{li}~\cite{park_extended_2002}; \\ \glsentryshort{me}~\cite{park_extended_2002}; \glsentryshort{stasch}~\cite{park_extended_2002}} & $\circ$ \\
        \midrule
        \glsentryshort{hdf}~\cite{girault_hierarchical_1999} & \glsentryshort{biso} & never & $\mathbb{N}^*$ & \makecell{\glsentryshort{hi} \\ \glsentryshort{it}} & \makecell{\glsentryshort{co}~\cite{girault_hierarchical_1999}; \glsentryshort{dec}~\cite{girault_hierarchical_1999}; \\ \glsentryshort{det}~\cite{girault_hierarchical_1999}; \glsentryshort{li}~\cite{girault_hierarchical_1999}} & $\circ$ \\
        \bottomrule
      \end{tabular}
      \begin{tablenotes}
        \item[1] only for the \enquote{SSDF actors}
        \item \emph{Rate and topology updates acronyms}: \glsentryshort{biso}: \glsentrylong{biso}, \glsentryshort{biro}: \glsentrylong{biro} \glsentryshort{wiso}: \glsentrylong{wiso}, \glsentryshort{wiro}: \glsentrylong{wiro}
        \item \emph{Features acronyms}: \glsentryshort{gs}: \glsentrylong{gs}, \glsentryshort{hi}: \glsentrylong{hi}, \glsentryshort{it}: \glsentrylong{it}, \glsentryshort{mdf}: \glsentrylong{mdf}, \glsentryshort{swi}: \glsentrylong{swi}
        \item \emph{Analyzability acronyms}: \glsentryshort{co}: \glsentrylong{co}, \glsentryshort{dec}: \glsentrylong{dec}, \glsentryshort{det}: \glsentrylong{det}, \glsentryshort{li}: \glsentrylong{li}, \glsentryshort{me}: \glsentrylong{me}, \glsentryshort{sco}: \glsentrylong{sco}, \glsentryshort{stasch}: \glsentrylong{stasch}
      \end{tablenotes}
    \end{threeparttable}
  \end{table*}
  
  \subsection{Phased-based DF MoCCs}

  \label{sec:phased-based-dfmoccs}

  This category includes \glspl{dfmocc} which have the \emph{phase} feature. This means that the number of tokens produced and consumed by an actor can vary from one job to another. Depending on the \gls{dfmocc}, the pattern of production and consumption can be cyclic or not and defined at compile-time or at \rt.

  \subsubsection{CSDF}

  The actors of a \gls{csdf}~\cite{bilsen_cyclostatic_1996} specification have cyclic execution function, production, and consumption rates which are established at compile-time. The production and consumption rates change periodically according to the defined cycle. The rates take their values in $\mathbb{N}$, so some channels may be periodically disabled when the rate is 0. The conversion from a \gls{csdf} specification to an \gls{hsdf} specification~\cite{bilsen_cyclostatic_1996} provides consistency and liveness checking, and reference~\cite{benazouz_liveness_2013} provides sufficient conditions to ensure the liveness without transformation into an \gls{hsdf} specification. Schedules can be derived at compile-time~\cite{anapalli_static_2009,bodin_periodic_2013}.

  \subsubsection{CDDF}

  \gls{cddf}~\cite{wauters_cyclodynamic_1996} is a dynamic version of \gls{csdf}. The execution function, token ratios, and execution sequence length can vary at \rt. A control token is read at each actor execution to determine its behavior. Restrictions are imposed on using the control token to enhance the analyzability. These restrictions permit to derive conditions about the strong consistency of a \gls{cddf} specification and its schedulability. These restrictions also ensure the functional determinism of \gls{cddf}.

  \subsubsection{PCG}

  \glspl{pcg}~\cite{bodin_fast_2014} extends \gls{csdf} and \glspl{cg} with both consumption thresholds and initialization phases. The rates of \glspl{pcg} are divided into initial and steady phases. The initial sequence is performed at the beginning of the execution, and the steady sequence, which is cyclic, takes over for the rest of the execution. The authors of~\cite{bodin_fast_2014} demonstrate conditions for consistency and liveness checkings and a lower bound for memory footprint.

  \subsubsection{FRDF}

  \gls{frdf}~\cite{oh_fractional_2004} is the first \gls{dfmocc} with fractional rates. A fractional rate $p/q$ guarantees that $p$ tokens are produced/consumed every $q$ executions. Consequently, some executions may produce/consume multiple tokens while some may not produce/consume any. The execution instances at which production and consumption occur are not fixed \eg for any three consecutive executions of an actor with a production rate of $1/3$, only one execution produces a token, and it may be the first \emph{or} the second \emph{or} the third, and this pattern may vary at \rt. Therefore, rates and topology updates are classified as \glsentryshort{wiro}. We will see that another \gls{dfmocc} called \polygraph~\cite{dubrulle_polygraph_2021}, which also has fractional rates, overcomes this limitation by using initial tokens. Interestingly, initial tokens are \emph{not} part of \gls{frdf}. Nothing prevents the token's production and consumption pattern of a fractional rate to depend on external factors. Thus, \gls{frdf} is non-deterministic. However, as a coarse-grained production and consumption pattern is guaranteed at compile-time by the fractional rates, a static schedule can be derived, and memory footprint minimization also follows.

  \begin{table*}[htbp]
    \centering
    \caption{Features and static analyses of \nameref{sec:phased-based-dfmoccs}.}
    \label{tab:phased-based-dfmoccs}
    \begin{threeparttable}[htbp]
      \begin{tabular}{|c|c|c|c|c|c|c|}
        \toprule
        \textbf{\glsentryshort{dfmocc}} & \makecell{\textbf{Rate} \\ \textbf{updates}} & \makecell{\textbf{Topology} \\ \textbf{updates}} & \makecell{\textbf{Range} \\ \textbf{rate}} & \textbf{Features} & \textbf{Static analyses} & \makecell{\textbf{Turing} \\ \textbf{complete}} \\
        \midrule
        \gls{csdf}~\cite{bilsen_cyclostatic_1996} & \glsentryshort{wiso} & \glsentryshort{wiso} & $\mathbb{N}$ & \makecell{\glsentryshort{it} \\ \glsentryshort{ph}} & \makecell{\glsentryshort{co}~\cite{bilsen_cyclostatic_1996}; \glsentryshort{dec}~\cite{anapalli_static_2009,bodin_periodic_2013}; \\ \glsentryshort{det}~\cite{bilsen_cyclostatic_1996}; \glsentryshort{li}~\cite{bilsen_cyclostatic_1996,benazouz_liveness_2013}; \\ \glsentryshort{stasch}~\cite{anapalli_static_2009,bodin_periodic_2013}} & $\circ$ \\
        \midrule
        \glsentryshort{cddf}~\cite{wauters_cyclodynamic_1996} & \glsentryshort{wiro} & \glsentryshort{wiro} & $\mathbb{N}$ & \makecell{\glsentryshort{it} \\ \glsentryshort{ph}} & \makecell{\glsentryshort{co}~\cite{wauters_cyclodynamic_1996}; \glsentryshort{dec}~\cite{wauters_cyclodynamic_1996}; \\ \glsentryshort{det}~\cite{wauters_cyclodynamic_1996}; \glsentryshort{sco}~\cite{wauters_cyclodynamic_1996}} & $\circ$ \\
        \midrule
        \gls{pcg}~\cite{bodin_fast_2014} & never & never & $\mathbb{N}$ & \makecell{\glsentryshort{ct} \\ \glsentryshort{inistep} \\ \glsentryshort{it} \\ \glsentryshort{ph}} & \makecell{\glsentryshort{co}~\cite{bodin_fast_2014}; \glsentryshort{dec}~\cite{bodin_fast_2014}; \\ \glsentryshort{det}~\cite{bodin_fast_2014}; \glsentryshort{li}~\cite{bodin_fast_2014}; \\ \glsentryshort{me}~\cite{bodin_fast_2014}} & $\circ$ \\
        \midrule
        \gls{frdf}~\cite{oh_fractional_2004} & \glsentryshort{wiro} & \glsentryshort{wiro} & $\mathbb{Q}^*$ & \glsentryshort{ph} & \makecell{\glsentryshort{co}~\cite{oh_fractional_2004}; \glsentryshort{dec}~\cite{oh_fractional_2004}; \\ \glsentryshort{li}~\cite{oh_fractional_2004}; \glsentryshort{me}~\cite{oh_fractional_2004}; \\ \glsentryshort{stasch}~\cite{oh_fractional_2004}} & $\circ$ \\
        \bottomrule
      \end{tabular}
      \begin{tablenotes}
        \item \emph{Rate and topology updates acronyms}: \glsentryshort{wiso}: \glsentrylong{wiso}, \glsentryshort{wiro}: \glsentrylong{wiro}
        \item \emph{Features acronyms}: \glsentryshort{ct}: \glsentrylong{ct}, \glsentryshort{inistep}: \glsentrylong{inistep}, \glsentryshort{it}: \glsentrylong{it}, \glsentryshort{ph}: \glsentrylong{ph}
        \item \emph{Analyzability acronyms}: \glsentryshort{co}: \glsentrylong{co}, \glsentryshort{dec}: \glsentrylong{dec}, \glsentryshort{det}: \glsentrylong{det}, \glsentryshort{li}: \glsentrylong{li}, \glsentryshort{me}: \glsentrylong{me}, \glsentryshort{sco}: \glsentrylong{sco}, \glsentryshort{stasch}: \glsentrylong{stasch}
      \end{tablenotes}
    \end{threeparttable}
  \end{table*}

  \subsection{Timed-based DF MoCCs}

  \label{sec:timed-based-dfmoccs}

  This category includes \glspl{dfmocc} where the concept of time is present. This can be the specification of execution time, a frequency execution, or even a delay.

  \subsubsection{tSDF}

  \gls{tsdf}~\cite{ghamarian_latency_2007} is an extension of \gls{sdf} with a function that maps each actor to their execution time. Such a function permits the extension of static analysis algorithms of \gls{sdf} to throughput and latency analyses. For instance, a trade-off between buffer requirements and throughput constraints is explored in~\cite{stuijk_exploring_2006}, and a linear programming formulation is proposed in~\cite{govindarajan_minimizing_2002} to compute the buffer size with optimal throughput \ie the maximum throughput without storage constraints. In reference~\cite{ghamarian_latency_2007} and~\cite{ghamarian_throughput_2006}, the authors propose a state-space traversal to find the minimum latency and maximum throughput of an \gls{sdf} specification, respectively.

  \subsubsection{tCSDF}

  \gls{tcsdf}~\cite{wiggers_efficient_2007} extends \gls{csdf} with a function that maps each actor's phase with a non-null execution time. Such a function permits the extension of static analysis of \gls{csdf} to throughput and latency analyses. Various studies have been conducted on the trade-off between throughput and memory footprint. The authors of~\cite{wiggers_efficient_2007} propose a heuristic algorithm to perform this trade-off, and reference~\cite{stuijk_throughputbuffering_2008} proposes a design-space exploration of the Pareto points in the throughput/buffer size space. \gls{milp} and \gls{mmlp} minimized buffer size under throughput constraints in~\cite{benazouz_new_2010} and~\cite{benazouz_cyclostatic_2013}, respectively. Throughput evaluation has also been studied using~\cite{bodin_optimal_2016}. Static schedules are also derived in reference~\cite{bodin_periodic_2013} while optimizing memory footprint~\cite{benazouz_cyclostatic_2013}.

  \subsubsection{CSDF$^a$}

  \gls{csdfa}~\cite{koek_csdfa_2016} extends \gls{csdf} by providing a mechanism to handle out-of-order token consumption. Mechanisms such as circular buffers or predefined buffers' access patterns limit the overhead induced by simultaneous executions of an actor. The authors explored the trade-off between the maximum concurrent executions and the memory footprint, throughput, and latency. \gls{csdfa} extends \gls{csdf}, so the static analysis algorithms of \gls{csdfa} apply to \gls{csdf} specification. However, it is unclear if the static analyses of \gls{csdf}, such as memory and static schedule, can be applied to an \gls{csdfa} specification.

  \subsubsection{TPDF}

  \gls{tpdf}~\cite{khanhdo_transaction_2016} extends \gls{csdf}. Rates can be parametric and updated between iterations. \gls{tpdf} also introduces three types of actors: \emph{select-duplicate}, \emph{transaction}, and \emph{clock} actors. \emph{Select-duplicate} replicates its single entry into any combinations of its outputs, \emph{transaction} is the symmetric process, and \emph{clock} actor sends a control token periodically to another actor. This control token defines the execution mode of the actor that consumes that control token \eg waiting for all input data to be available before execution or selecting the data with the highest priority. Under some restrictions, two different execution modes can be used within the same iteration of a \gls{tpdf} specification, thus allowing topology and some rate updates within iterations. \gls{tpdf} provides consistency and liveness checking, as well as a scheduling strategy.

  \subsubsection{PolyGraph}

  \polygraph~\cite{dubrulle_data_2019} enhances the semantics of rational rates of \gls{frdf}. A rate of $p/q$ means $p$ tokens are produced/consumed every $q$ executions. An actor's execution increases/decreases by $\frac{p}{q}$ the fractional number of tokens in the channels involved. In contrast with \gls{frdf}, initials tokens are used to derive a unique execution sequence from a rational rate. An actor may also have a frequency constraint and a delay. Thus, it must execute at that frequency, and its first execution occurs after the delay. A \polygraph specification is statically analyzable in terms of consistency and liveness. Reference~\cite{hamelin_performance_2024} also provides execution windows derivation and a schedulability test.

  \subsubsection{Dynamic PolyGraph}

  Reference~\cite{dubrulle_dynamic_2019} presents a dynamic extension of \polygraph. Actors of dynamic \polygraph label tokens with an execution mode. The execution mode an actor consumes defines how those tokens are processed. A peculiarity of dynamic \polygraph is that tokens can be produced with empty content \ie they do not carry any meaningful data. Actors consuming such tokens consider that the corresponding input channel is (virtually) disabled. An implicit assumption is that channels must always be active to transit tokens, even if those tokens have empty content. Nevertheless, dynamic \polygraph is statically analyzable in terms of its consistency and liveness.

  \subsubsection{ppSDF}

  \gls{ppsdf}~\cite{honorat_scheduling_2020} extends \gls{sdf} by allowing a subset of actors to have a frequency constraint but not a phase as in \polygraph. The authors give some conditions to assert the schedulability of a \gls{ppsdf} specification.

  \subsubsection{VSDF}

  \gls{vsdf}~\cite{kerihuel_vsdf_1994} extends \gls{sdf} for specifying \gls{vlsi} systems. The authors augment the consistency analysis of \gls{sdf} by including temporal constraints. This analysis includes equations to formalize that tokens produced on a channel at a given time instant are also consumed at this same time instant. Therefore, not only the number of tokens produced and consumed on each channel must be equal, but also the time at which the tokens are produced and consumed must be equal. It is more accurate to specify the timing constraints on production and consumption instant instead of on actors. Indeed, when timing constraints are specified on the actor, the production and consumption instant are within the frequency interval but have no fixed value.

  \subsubsection{HSDF$^a$}

  \gls{hsdfa}~\cite{kuiper_latency_2017} extends \gls{hsdf} by determining the consumption order of tokens with static indices independently of the production order. The token produced by the $(n+1)$-th job may precede the token produced by the $n$-th job in the buffer's channel if the $n$-th job finishes before the $(n+1)$-th job. The authors of~\cite{kuiper_latency_2017} propose a method to compute the end-to-end latency of \gls{hsdfa} specifications. To that end, assigning an execution time is essential. They suggest deriving a timed automaton~\cite{alur_theory_1994} semantically equivalent to an \gls{hsdfa} specification and using a model-checker like \glsentryshort{uppaal}~\cite{david_modelbased_2009} to compute the exact end-to-end latency of the initial \gls{hsdfa} specification. In addition, as \gls{hsdfa} is a restriction of \gls{csdfa}~\cite{koek_csdfa_2016}, it inherits its analyzability.

  \subsubsection{ILDF}

  \gls{ildf}~\cite{teich_analysis_2006} extends \gls{sdf} by permitting consumption and production rates to be within a finite natural integers interval. The actual rates are determined just before the start of the execution and remain fixed. A schedule and the worst-case memory footprint can be derived at compile-time. In addition, if the actors' \gls{bcet} and \gls{wcet} are known, the schedule's best-case and worst-case time performance can be computed. Determining the fixed rate before the execution depends on external conditions \eg the size of the data to be processed. Therefore, \gls{ildf} is not deterministic.

  \subsubsection{VRDF}

  \gls{vrdf}~\cite{wiggers_buffer_2008} is a parametric extension of \gls{sdf} that imposes constraints on parameter usage. For instance, a parameter is used by at most two actors, namely the modifier and user. This means that a parameter can be set at each execution of the modifier, even within an iteration of a \gls{vrdf} specification. The process of determining the new parameter values is not explicitly defined, and it could result from non-determinism procedures. We also assume that the assignment parameter process does not depend on \rt conditions. In a \gls{vrdf} specification, two actors using the same parameter must have the same entry in the repetition vector. These and other restrictions allow the authors to prove the strong consistency of a \gls{vrdf} specification. They propose another static analysis to determine the parameters assignation that guarantees the fulfillment of a throughput constraint of a \gls{vrdf} specification, as well as the associated memory footprint.

  \subsubsection{VPDF}

  \gls{vpdf}~\cite{wiggers_buffer_2010} extends \gls{vrdf}. Besides structural constraints similar to the one of \gls{vrdf}, actors of a \gls{vpdf} specification have two parameters for each phase: the number of repetitions and the rate of that phase.

  \subsubsection{RMDF}

  \gls{rmdf}~\cite{roumage_realtime_2025} builds upon \polygraph by enabling the specification and analysis of \glspl{cps} with both timing constraints on some actors and a mode-dependent execution \ie an execution with conditional execution branches. An \gls{rmdf} specification is statically analyzable regarding its consistency and liveness, and the execution window length can be derived to facilitate a feasibility test \cite{roumage_realtime_2025}.

  \begin{table*}[htbp]
    \centering
    \caption{Features and static analyses of Timed-based DF MoCCs.}
    \label{tab:timed-based-dfmoccs}
    \begin{threeparttable}[htbp]
      \begin{tabular}{|c|c|c|c|c|c|c|}
        \toprule
        \textbf{\glsentryshort{dfmocc}} & \makecell{\textbf{Rate} \\ \textbf{updates}} & \makecell{\textbf{Topology} \\ \textbf{updates}} & \makecell{\textbf{Range} \\ \textbf{rate}} & \textbf{Features} & \textbf{Static analyses} & \makecell{\textbf{Turing} \\ \textbf{complete}} \\
        \midrule
        \glsentryshort{tsdf}~\cite{ghamarian_latency_2007} & never & never & $\mathbb{N}^*$ & \makecell{\glsentryshort{et} \\ \glsentryshort{it}} & \makecell{\glsentryshort{co}~\cite{govindarajan_minimizing_2002,ghamarian_latency_2007}; \glsentryshort{dec}~\cite{govindarajan_minimizing_2002,ghamarian_latency_2007}; \\ \glsentryshort{det}~\cite{bilsen_cyclostatic_1996}; \glsentryshort{la}~\cite{ghamarian_latency_2007}; \\ \glsentryshort{li}~\cite{govindarajan_minimizing_2002,ghamarian_latency_2007}; \glsentryshort{me}~\cite{govindarajan_minimizing_2002, stuijk_exploring_2006}; \\ \glsentryshort{stasch}~\cite{govindarajan_minimizing_2002,ghamarian_latency_2007}; \\ \glsentryshort{th}~\cite{govindarajan_minimizing_2002, stuijk_exploring_2006,ghamarian_throughput_2006}} & $\circ$ \\
        \midrule
        \glsentryshort{tcsdf}~\cite{wiggers_efficient_2007} & \glsentryshort{wiso} & \glsentryshort{wiso} & $\mathbb{N}$ & \makecell{\glsentryshort{et} \\ \glsentryshort{it} \\ \glsentryshort{ph}} & \makecell{\glsentryshort{co}~\cite{bodin_periodic_2013,benazouz_cyclostatic_2013}; \glsentryshort{dec}~\cite{wiggers_efficient_2007}; \\ \glsentryshort{det}~\cite{wiggers_efficient_2007}; \glsentryshort{li}~\cite{bodin_periodic_2013,benazouz_cyclostatic_2013}; \\ \glsentryshort{me}~\cite{wiggers_efficient_2007,stuijk_throughputbuffering_2008,benazouz_new_2010,benazouz_cyclostatic_2013}; \\ \glsentryshort{stasch}~\cite{bodin_periodic_2013,benazouz_cyclostatic_2013,bodin_optimal_2016}; \\  \glsentryshort{th}~\cite{wiggers_efficient_2007,stuijk_throughputbuffering_2008,benazouz_new_2010,benazouz_cyclostatic_2013,bodin_optimal_2016}} & $\circ$ \\
        \midrule
        \glsentryshort{csdfa}~\cite{koek_csdfa_2016} & \glsentryshort{wiso} & \glsentryshort{wiso} & $\mathbb{N}$ & \makecell{\glsentryshort{et} \\ \glsentryshort{it} \\ \glsentryshort{ooc} \\ \glsentryshort{ph}} & \makecell{\glsentryshort{co}~\cite{bilsen_cyclostatic_1996}; \glsentryshort{dec}~\cite{koek_csdfa_2016}; \\ \glsentryshort{det}~\cite{koek_csdfa_2016}; \glsentryshort{la}~\cite{koek_csdfa_2016}; \\ \glsentryshort{li}~\cite{bilsen_cyclostatic_1996}; \glsentryshort{me}~\cite{koek_csdfa_2016}; \\ \glsentryshort{th}~\cite{koek_csdfa_2016}} & $\circ$ \\
        \midrule
        \gls{tpdf}~\cite{khanhdo_transaction_2016} & \makecell{\glsentryshort{wiso} + \\ \glsentryshort{wiro}} & \makecell{\glsentryshort{wiso} + \\ \glsentryshort{wiro}} & $\mathbb{N}$ & \makecell{\glsentryshort{et} \\ \glsentryshort{freq} \\ \glsentryshort{it} \\ \glsentryshort{pa} \\ \glsentryshort{ph}} & \makecell{\glsentryshort{co}~\cite{khanhdo_transaction_2016}; \glsentryshort{dec}~\cite{khanhdo_transaction_2016}; \\ \glsentryshort{det}~\cite{khanhdo_transaction_2016}; \glsentryshort{li}~\cite{khanhdo_transaction_2016}; \\ \glsentryshort{stasch}~\cite{khanhdo_transaction_2016}; \glsentryshort{th}~\cite{khanhdo_model_2016}} & $\circ$ \\
        \midrule
        \polygraph~\cite{dubrulle_polygraph_2021} & \glsentryshort{wiso} & \glsentryshort{wiso} & $\mathbb{Q}^*$ & \makecell{\glsentryshort{del} \\ \glsentryshort{et} \\ \glsentryshort{freq} \\ \glsentryshort{it} \\ \glsentryshort{ph}} & \makecell{\glsentryshort{co}~\cite{dubrulle_polygraph_2021}; \glsentryshort{dec}~\cite{hamelin_performance_2024}; \\ \glsentryshort{det}~\cite{dubrulle_polygraph_2021}; \glsentryshort{execwin}~\cite{hamelin_performance_2024,roumage_static_2024}; \\ \glsentryshort{li}~\cite{dubrulle_polygraph_2021}; \glsentryshort{stasch}~\cite{hamelin_performance_2024}} & $\circ$ \\
        \midrule
        \makecell{Dynamic \\ \polygraph~\cite{dubrulle_dynamic_2019}} & \glsentryshort{wiso} & \glsentryshort{wiso} & $\mathbb{Q}^*$ & \makecell{\glsentryshort{del} \\ \glsentryshort{freq} \\ \glsentryshort{it} \\ \glsentryshort{ph}} & \makecell{\glsentryshort{co}~\cite{dubrulle_dynamic_2019}; \glsentryshort{dec}~\cite{dubrulle_dynamic_2019}; \\ \glsentryshort{det}~\cite{dubrulle_dynamic_2019}; \glsentryshort{li}~\cite{dubrulle_dynamic_2019}} & $\circ$ \\
        \midrule
        \gls{ppsdf}~\cite{honorat_scheduling_2020} & never & never & $\mathbb{N}^*$ & \makecell{\glsentryshort{et} \\ \glsentryshort{freq} \\ \glsentryshort{it}} & \makecell{\glsentryshort{co}~\cite{honorat_scheduling_2020}; \glsentryshort{dec}~\cite{honorat_scheduling_2020}; \\ \glsentryshort{det}~\cite{honorat_scheduling_2020}; \glsentryshort{li}~\cite{honorat_scheduling_2020}; \\ \glsentryshort{stasch}~\cite{honorat_scheduling_2020}} & $\circ$ \\
        \midrule
        \gls{vsdf}~\cite{kerihuel_vsdf_1994} & never & never & $\mathbb{N}^*$ & \makecell{\glsentryshort{it} \\ \glsentryshort{pci}} & \makecell{\glsentryshort{co}~\cite{kerihuel_vsdf_1994}; \glsentryshort{dec}~\cite{kerihuel_vsdf_1994}; \\ \glsentryshort{det}~\cite{kerihuel_vsdf_1994}; \glsentryshort{li}~\cite{kerihuel_vsdf_1994}} & $\circ$ \\
        \midrule
        \gls{hsdfa}~\cite{kuiper_latency_2017} & never & never & $\{1\}$ & \makecell{\glsentryshort{et} \\ \glsentryshort{it} \\ \glsentryshort{ooc}} & \makecell{\glsentryshort{co}~\cite{koek_csdfa_2016}; \glsentryshort{dec}~\cite{kuiper_latency_2017}; \\ \glsentryshort{det}~\cite{kuiper_latency_2017}; \glsentryshort{la}~\cite{koek_csdfa_2016}; \\ \glsentryshort{li}~\cite{kuiper_latency_2017}} & $\circ$ \\
        \midrule
        \gls{ildf}~\cite{teich_analysis_2006} & never & never & $\mathbb{N}^*$ & \makecell{\glsentryshort{et} \\ \glsentryshort{it} \\ \glsentryshort{rai}} & \makecell{\glsentryshort{co}~\cite{teich_analysis_2006}; \glsentryshort{dec}~\cite{teich_analysis_2006}; \\ \glsentryshort{det}~\cite{teich_analysis_2006}; \glsentryshort{me}~\cite{teich_analysis_2006}; \\ \glsentryshort{la}~\cite{teich_analysis_2006}; \glsentryshort{li}~\cite{teich_analysis_2006}; \\ \glsentryshort{stasch}~\cite{teich_analysis_2006}} & $\circ$ \\
        \midrule
        \glsentryshort{vrdf}~\cite{wiggers_buffer_2008} & \glsentryshort{wiso} & \glsentryshort{wiso} & $\mathbb{N}$ & \makecell{\glsentryshort{et} \\ \glsentryshort{it} \\ \glsentryshort{pa}} & \makecell{\glsentryshort{co}~\cite{wiggers_buffer_2008}; \glsentryshort{sco}~\cite{wiggers_buffer_2008}; \\ \glsentryshort{th}~\cite{wiggers_buffer_2008}; \glsentryshort{me}~\cite{wiggers_buffer_2008}} & $\circ$ \\
        \midrule
        \glsentryshort{vpdf}~\cite{wiggers_buffer_2010} & \glsentryshort{biso} & \glsentryshort{biso} & $\mathbb{N}$ & \makecell{\glsentryshort{et} \\ \glsentryshort{it} \\ \glsentryshort{pa} \\ \glsentryshort{ph}} & \makecell{\glsentryshort{co}~\cite{wiggers_buffer_2010}; \glsentryshort{sco}~\cite{wiggers_buffer_2010}; \\ \glsentryshort{th}~\cite{wiggers_buffer_2010}; \glsentryshort{me}~\cite{wiggers_buffer_2010}} & $\circ$ \\
        \midrule
        \glsentryshort{rmdf}~\cite{roumage_realtime_2025} & \makecell{\glsentryshort{wiso} + \\ \glsentryshort{wiro}} & \makecell{\glsentryshort{wiso} + \\\glsentryshort{wiro}} & $\mathbb{Q}^*$ & \makecell{\glsentryshort{del} \\ \glsentryshort{et} \\ \glsentryshort{freq} \\ \glsentryshort{it} \\ \glsentryshort{pa} \\ \glsentryshort{ph}} & \makecell{\glsentryshort{co}~\cite{roumage_realtime_2025}; \glsentryshort{dec}~\cite{roumage_realtime_2025}; \\ \glsentryshort{execwin}~\cite{roumage_realtime_2025}; \\ \glsentryshort{det}~\cite{roumage_realtime_2025}; \glsentryshort{li}~\cite{roumage_realtime_2025}} & $\circ$ \\
        \bottomrule
      \end{tabular}
      \begin{tablenotes}
        \item \emph{Rate and topology updates acronym}: \glsentryshort{biso}: \glsentrylong{biso}, \glsentryshort{wiso}: \glsentrylong{wiso}, \glsentryshort{wiro}: \glsentrylong{wiro}
        \item \emph{Features acronyms}: \glsentryshort{del}: \glsentrylong{del}, \glsentryshort{et}: \glsentrylong{et}, \glsentryshort{freq}: \glsentrylong{freq}, \glsentryshort{it}: \glsentrylong{it}, \glsentryshort{ooc}: \glsentrylong{ooc}, \glsentryshort{pa}: \glsentrylong{pa}, \glsentryshort{pci}: \glsentrylong{pci}, \glsentryshort{ph}: \glsentrylong{ph}, \glsentryshort{rai}: \glsentrylong{rai}
        \item \emph{Analyzability acronyms}: \glsentryshort{co}: \glsentrylong{co}, \glsentryshort{dec}: \glsentrylong{dec}, \glsentryshort{det}: \glsentrylong{det}, \glsentryshort{la}: \glsentrylong{la}, \glsentryshort{li}: \glsentrylong{li}, \glsentryshort{stasch}: \glsentrylong{stasch}, \glsentryshort{sco}: \glsentrylong{sco}
      \end{tablenotes}
    \end{threeparttable}
  \end{table*}

  \subsection{Boolean-based DF MoCCs}

  \label{sec:boolean-based-dfmoccs}

  This category includes \glspl{dfmocc} where parameters taking their value between 0 and 1 are used to control actors' execution. Those values determine either how actors consume or produce tokens or determine the topology of the \gls{dfg}.

  \subsubsection{BDF}

  \gls{bdf}~\cite{buck_scheduling_1993-1} is one of the first \gls{dfmocc} focusing on topological updates. The production/consumption rates of specific actors called \emph{switch}/\emph{select} are either 0 or 1, and control tokens consumed by those actors determine which port is used. The control tokens enforce a data-dependency behavior. \gls{bdf} is Turing complete, and this latter result is often used to demonstrate the Turing completeness of other \glspl{dfmocc}. The extension of static analyses of \gls{sdf} to \gls{bdf} is discussed in~\cite{buck_scheduling_1993-1}.

  \subsubsection{IDF}

  \gls{idf}~\cite{buck_scheduling_1993-1} is a generalization of \gls{bdf} where control tokens are any integer. Thus, the switch and select actors become \emph{case} and \emph{end-case} actors. Some static analysis algorithms of \gls{bdf} can be extended to \gls{idf}.

  \subsubsection{BPDF}

  \gls{bpdf}~\cite{bebelis_bpdf_2013} combines two parameters: integer parameters express dynamic rates and 2-values parameters (i.e., \enquote{boolean parameters}) on the channels. Those latter dynamically (des)activate the channels, possibly within iterations of a \gls{bpdf} specification. In order to preserve the consistency and liveness analysis, boolean parameters are allowed to change at some well-defined points in the executions. An approach to schedule a \gls{bpdf} specification is also proposed. There is no restriction on how the value of boolean parameters is chosen, so \gls{bpdf} is not deterministic.

  \begin{table*}[htbp]
    \centering
    \caption{Features and static analyses of \nameref{sec:boolean-based-dfmoccs}.}
    \label{tab:boolean-based-dfmoccs}
    \begin{threeparttable}[htbp]
      \begin{tabular}{|c|c|c|c|c|c|c|}
        \toprule
        \textbf{\glsentryshort{dfmocc}} & \makecell{\textbf{Rate} \\ \textbf{updates}} & \makecell{\textbf{Topology} \\ \textbf{updates}} & \makecell{\textbf{Range} \\ \textbf{rate}} & \textbf{Features} & \textbf{Static analyses} & \makecell{\textbf{Turing} \\ \textbf{complete}} \\
        \midrule
        \gls{bdf}~\cite{buck_scheduling_1993-1} & \glsentryshort{wiso} & \glsentryshort{wiso} & $\mathbb{N}$ & \glsentryshort{it} & N/A\tnote{1} & $\bullet$ \\
        \midrule
        \gls{idf}~\cite{buck_scheduling_1993-1} & \glsentryshort{wiso} & \glsentryshort{wiso} & $\mathbb{N}$ & \glsentryshort{it} & N/A\tnote{1} & $\bullet$ \\
        \midrule
        \gls{bpdf}~\cite{bebelis_bpdf_2013} & \glsentryshort{wiso} & \glsentryshort{wiso} & $\mathbb{N}$ & \makecell{\glsentryshort{it} \\ \glsentryshort{pa}} & \makecell{\glsentryshort{co}~\cite{bebelis_bpdf_2013}; \glsentryshort{dec}~\cite{bebelis_bpdf_2013}; \\ \glsentryshort{li}~\cite{bebelis_bpdf_2013}; \glsentryshort{stasch}~\cite{bebelis_bpdf_2013}} & $\circ$ \\
        \bottomrule
      \end{tabular}
      \begin{tablenotes}
        \item[1] a subclass of \gls{bdf} and \gls{idf} models are analyzable
        \item \emph{Rate and topology updates acronyms}: \glsentryshort{wiso}: \glsentrylong{wiso}
        \item \emph{Features acronyms}: \glsentryshort{it}: \glsentrylong{it}, \glsentryshort{pa}: \glsentrylong{pa}
        \item \emph{Analyzability acronyms}: \glsentryshort{co}: \glsentrylong{co}, \glsentryshort{dec}: \glsentrylong{dec}, \glsentryshort{li}: \glsentrylong{li}, \glsentryshort{stasch}: \glsentrylong{stasch}
      \end{tablenotes}
    \end{threeparttable}
  \end{table*}

  \subsection{Scenario-based DF MoCCs}

  \label{sec:scenario-based-dfmoccs}

  This category includes \glspl{dfmocc} where the execution of the actors is driven by scenarios. All \glspl{dfmocc} of this category are derived from or an extension of \gls{sadf}.

  \subsubsection{SADF}

  A system specified with \gls{sadf}~\cite{theelen_scenarioaware_2008} has a set of \emph{scenarios}. Besides assigning values to parametric rates, a scenario determines the execution times of the actors of an \gls{sadf} specification. Execution times are chosen from discrete and finite-support probability distributions, which is a key difference from \gls{esadf}, which will be described further in the paper. \emph{Detectors} are a special type of \gls{sadf} actors that model the control part of the system by dynamically detecting scenarios. The choice of the scenario may be non-deterministic. An \gls{sadf} specification switches arbitrarily between scenarios -even within an iteration of the \gls{sadf} specification. In such cases, there are \emph{subscenarios} within a scenario. Consequently, a subscenario change in the middle of an iteration may lead to meaningless behavior, and the authors of~\cite{theelen_scenarioaware_2008} have defined the strong consistency property of an \gls{sadf} specification, which ensures each detector executes only once per scenario. Besides a strong consistency analysis, a liveness checking and quantitative evaluation of memory footprint, throughput, and latency have been conducted~\cite{theelen_performance_2011}.

  \subsubsection{FSM-SADF}

  \gls{fsm-sadf}~\cite{stuijk_fsmbased_2008} is a restriction of \gls{sadf}. As in an \gls{sadf} specification, the dynamic behavior of a system is also viewed as an evolving sequence of static behaviors specified with an \gls{sdf} specification. However, scenarios can change only between iterations of an \gls{sdf} specification of the respective scenario and there are no subscenarios. A non-deterministic \gls{fsm} specifies the order in which the scenarios occur and the rates of the current scenario. Techniques to compute worst-case throughput and latency are discussed in~\cite{stuijk_scenarioaware_2011}. An exact computation is provided in~\cite{geilen_worstcase_2010}, while a trade-off between buffer size and throughput is explored in~\cite{ara_throughputbuffering_2018}. An \gls{fsm-sadf} specification is an \gls{sadf} specification that is strongly consistent. In contrast with \gls{sadf}, where actors' execution times follow a probability law, the execution times of actors in \gls{fsm-sadf} are fixed.

  \subsubsection{FSM-PSADF}

  \gls{fsm-psadf}~\cite{skelin_parameterized_2017} enhances \gls{fsm-sadf} by using parameters to improve the compactness. The scenario and its parameter configuration are both non-deterministically chosen at the end of an iteration. Reference~\cite{skelin_parameterized_2017} developed a state-space analysis to derive throughput and latency.

  \subsubsection{eSADF}

  \gls{esadf}~\cite{katoen_exponentially_2014} is a variant of \gls{sadf} where negative exponential distributions govern the execution times of actors. This assumption permits the use of \glspl{ma}~\cite{deng_semantics_2013} to capture the semantics of \gls{esadf}. Analysis techniques of \gls{ma}~\cite{guck_modelling_2013} yield a quantitative evaluation of memory footprint, throughput, or latency. Other metrics, such as the probability distribution of tokens in a channel, can also be computed. We assume that the authors of~\cite{katoen_exponentially_2014} consider only consistent and live \gls{esadf} specifications. The consistency and liveness can be checked using the same techniques as \gls{sadf} since those do not depend on the execution time of the actors.

  \subsubsection{xSADF}

  \gls{xsadf}~\cite{hartmanns_flexible_2016} extends both \gls{sadf} and \gls{esadf}. Besides supporting actors' execution times that follow arbitrary probability distributions, \gls{xsadf} endows actors with an additional cost function \eg energy usage, to have a finer analysis of the system's performance. Another extension is the process of selecting the scenario. In \gls{sadf} and \gls{esadf}, scenarios follow a probabilistic distribution known at compile-time. \gls{xsadf} relaxes this assumption and allows this choice to depend on external factors and possibly depend on \rt condition. This is why we consider that the rate and update topology of \gls{xsadf} is \glsentryshort{wiro} instead of \glsentryshort{wiso} as for \gls{sadf} and \gls{esadf}.

  \begin{table*}[htbp]
    \centering
    \caption{Features and static analyses of \nameref{sec:scenario-based-dfmoccs}.}
    \label{tab:scenario-based-dfmoccs}
    \begin{threeparttable}[htbp]
      \begin{tabular}{|c|c|c|c|c|c|c|}
        \toprule
        \textbf{\glsentryshort{dfmocc}} & \makecell{\textbf{Rate} \\ \textbf{updates}} & \makecell{\textbf{Topology} \\ \textbf{updates}} & \makecell{\textbf{Range} \\ \textbf{rate}} & \textbf{Features} & \textbf{Static analyses} & \makecell{\textbf{Turing} \\ \textbf{complete}} \\
        \midrule
        \gls{sadf}~\cite{theelen_scenarioaware_2008} & \glsentryshort{wiso} & \glsentryshort{wiso} & $\mathbb{N}$ & \makecell{\glsentryshort{et} \\ \glsentryshort{it} \\ \glsentryshort{pa}} & \makecell{\glsentryshort{co}~\cite{theelen_scenarioaware_2008}; \glsentryshort{dec}~\cite{theelen_scenarioaware_2008}; \\ \glsentryshort{la}~\cite{theelen_performance_2011}; \glsentryshort{li}~\cite{theelen_scenarioaware_2008}; \\ \glsentryshort{me}~\cite{theelen_performance_2011}; \glsentryshort{sco}~\cite{theelen_scenarioaware_2008}; \\ \glsentryshort{th}~\cite{theelen_performance_2011,geilen_performance_2017}} & $\circ$ \\
        \midrule
        \gls{fsm-sadf}~\cite{stuijk_fsmbased_2008} & \glsentryshort{biso} & \glsentryshort{biso} & $\mathbb{N}^*$ & \makecell{\glsentryshort{et} \\ \glsentryshort{it}} & \makecell{\glsentryshort{co}~\cite{stuijk_fsmbased_2008}; \glsentryshort{dec}~\cite{stuijk_fsmbased_2008}; \\ \glsentryshort{me}~\cite{ara_throughputbuffering_2018}; \glsentryshort{la}~\cite{stuijk_scenarioaware_2011}; \\ \glsentryshort{li}~\cite{stuijk_fsmbased_2008}; \glsentryshort{th}~\cite{geilen_worstcase_2010,stuijk_scenarioaware_2011,ara_throughputbuffering_2018}} & $\circ$ \\
        \midrule
        \gls{fsm-psadf}~\cite{skelin_parameterized_2017} & \glsentryshort{biso} & \glsentryshort{biso} & $\mathbb{N}^*$ & \makecell{\glsentryshort{et} \\ \glsentryshort{it} \\ \glsentryshort{pa}} & \makecell{\glsentryshort{co}~\cite{skelin_parameterized_2017}; \glsentryshort{dec}~\cite{skelin_parameterized_2017}; \\ \glsentryshort{la}~\cite{skelin_parameterized_2017}; \glsentryshort{li}~\cite{skelin_parameterized_2017}; \\ \glsentryshort{qsc}~\cite{skelin_parameterized_2017}; \glsentryshort{th}~\cite{skelin_parameterized_2017}} & $\circ$ \\
        \midrule
        \gls{esadf}~\cite{katoen_exponentially_2014} & \glsentryshort{wiso} & \glsentryshort{wiso} & $\mathbb{N}$ & \makecell{\glsentryshort{et} \\ \glsentryshort{it} \\ \glsentryshort{pa}} & \makecell{\glsentryshort{co}~\cite{theelen_scenarioaware_2008}; \glsentryshort{dec}~\cite{theelen_scenarioaware_2008}; \\ \glsentryshort{la}~\cite{katoen_exponentially_2014}; \glsentryshort{li}~\cite{theelen_scenarioaware_2008}; \\ \glsentryshort{me}~\cite{katoen_exponentially_2014}; \glsentryshort{sco}~\cite{theelen_scenarioaware_2008}; \\ \glsentryshort{th}~\cite{katoen_exponentially_2014}} & $\circ$ \\
        \midrule
        \gls{xsadf}~\cite{hartmanns_flexible_2016} & \glsentryshort{wiro} & \glsentryshort{wiro} & $\mathbb{N}$ & \makecell{\glsentryshort{et} \\ \glsentryshort{it} \\ \glsentryshort{pa}} & \makecell{\glsentryshort{la}~\cite{hartmanns_flexible_2016}; \glsentryshort{me}~\cite{hartmanns_flexible_2016}; \\ \glsentryshort{th}~\cite{hartmanns_flexible_2016}} & $\circ$ \\
        \bottomrule
      \end{tabular}
      \begin{tablenotes}
        \item \emph{Rate and topology updates acronyms}: \glsentryshort{biso}: \glsentrylong{biso}, \glsentryshort{wiso}: \glsentrylong{wiso}, \glsentryshort{wiro}: \glsentrylong{wiro}
        \item \emph{Features acronyms}: \glsentryshort{et}: \glsentrylong{et}, \glsentryshort{it}: \glsentrylong{it}, \glsentryshort{pa}: \glsentrylong{pa}
        \item \emph{Analyzability acronyms}: \glsentryshort{co}: \glsentrylong{co}, \glsentryshort{dec}: \glsentrylong{dec}, \glsentryshort{la}: \glsentrylong{la}, \glsentryshort{li}: \glsentrylong{li}, \glsentryshort{me}: \glsentrylong{me}, \glsentryshort{sco}: \glsentrylong{sco}, \glsentryshort{th}: \glsentrylong{th}
      \end{tablenotes}
    \end{threeparttable}
  \end{table*}

  \subsection{Meta-Models DF MoCCs}

  \label{sec:meta-model-dfmoccs}

  This category includes \glspl{dfmocc} which can be applied on the top of another \gls{dfmocc} to extend its expressiveness.

  \subsubsection{PSDF}

  \gls{psdf}~\cite{bhattacharya_parameterized_2001} is a parametric meta-model applied to \gls{sdf}. An actor of a \gls{psdf} specification is either \emph{primitive} or \emph{hierarchical}. A primitive one is composed of three graphs: the \emph{init}, \emph{subinit} and \emph{body} graphs. The body graph models the actor's behavior, and the init and subinit ones handle parameter reconfiguration. In order to maintain a valuable level of predictability, some parameter updates are restricted to occur at the boundaries of an iteration of the \gls{psdf} specification. The init graph handles such parameter updates, while the other is left to the subinit graph. An actor is hierarchical if its body graph is itself a \gls{psdf} specification. The authors of \gls{psdf} propose a quasi-static scheduling technique for acyclic \glspl{psdf} specification.

  \subsubsection{PCSDF}

  The authors of \gls{psdf}~\cite{bhattacharya_parameterized_2001} also apply their method to \gls{csdf} and yield the \gls{pcsdf}. The parameterization of \gls{pcsdf} is less expressive than \gls{vpdf}: phases' ratios and sequence execution length are parameterized, while in \gls{vpdf}, an additional parameter to each phase permits repetition a parametric number of times.

  \subsubsection{HPDF}

  \gls{hpdf}~\cite{sen_dataflowbased_2007} is a \gls{dfmocc} that refines a top-level actor of the \gls{hpdf} specification using any \gls{dfmocc} with a well-defined notion of iteration \eg \gls{sdf}, \gls{csdf}, or \gls{mdsdf}. However, from our understanding of \gls{hpdf}, top-level actors cannot be refined into a \gls{dfmocc} with timing constraints. There can be initial tokens between two actors of an \gls{hpdf} specification. An \gls{hpdf} specification executes in bounded memory if its actors execute in bounded memory.

  \subsubsection{PIMM}

  \gls{pimm}~\cite{desnos_pimm_2013} extends the semantics of any deterministic \gls{dfmocc}. The \gls{pimm} model uses an interface-based hierarchy and a set of parameters. Applying \gls{pimm} to \gls{sdf} yields the \gls{pisdf}, which can be seen as an extension of \gls{ibsdf}.

  \subsubsection{SAD}

  \gls{sad}~\cite{arrestier_delays_2018} tackles the memory persistence of initial tokens across the \gls{sad}'s model iterations. \gls{sad} extends the semantics of the initial tokens with an explicit initialization/discard at the start/end of each iteration.

  \begin{table*}[htbp]
    \centering
    \caption{Features and static analyses of \nameref{sec:meta-model-dfmoccs}.}
    \label{tab:meta-model-dfmoccs}
    \begin{threeparttable}[htbp]
      \begin{tabular}{|c|c|c|c|c|c|c|}
        \toprule
        \textbf{\glsentryshort{dfmocc}} & \makecell{\textbf{Rate} \\ \textbf{updates}} & \makecell{\textbf{Topology} \\ \textbf{updates}} & \makecell{\textbf{Range} \\ \textbf{rate}} & \textbf{Features} & \textbf{Static analyses} & \makecell{\textbf{Turing} \\ \textbf{complete}} \\
        \midrule
        \gls{psdf}~\cite{bhattacharya_parameterized_2001} & \glsentryshort{wiso} & never & $\mathbb{N}^*$ & \makecell{\glsentryshort{hi} \\ \glsentryshort{mm} \\ \glsentryshort{pa}} & \makecell{\glsentryshort{co}\tnote{1}~\cite{bhattacharya_parameterized_2001}; \glsentryshort{li}\tnote{1}~\cite{bhattacharya_parameterized_2001}; \\ \glsentryshort{qsc}\tnote{1}~\cite{bhattacharya_parameterized_2001}} & N/A\tnote{3} \\
        \midrule
        \glsentryshort{pcsdf}~\cite{bhattacharya_parameterized_2001} & \glsentryshort{wiso} & never & $\mathbb{N}$ & \makecell{\glsentryshort{hi} \\ \glsentryshort{mm} \\ \glsentryshort{pa} \\ \glsentryshort{ph}} & \makecell{\glsentryshort{co}\tnote{2}~\cite{bhattacharya_parameterized_2001}; \glsentryshort{li}\tnote{2}~\cite{bhattacharya_parameterized_2001}; \\ \glsentryshort{qsc}\tnote{2}~\cite{bhattacharya_parameterized_2001}} & N/A\tnote{3} \\
        \midrule
        \gls{hpdf}~\cite{sen_dataflowbased_2007} & \glsentryshort{biso} & never & $\mathbb{N}^*$ & \makecell{\glsentryshort{mm} \\ \glsentryshort{pa}} & \makecell{\glsentryshort{co}~\cite{sen_dataflowbased_2007}; \glsentryshort{li}~\cite{sen_dataflowbased_2007}; \\ \glsentryshort{qsc}~\cite{sen_dataflowbased_2007}} & N/A\tnote{3} \\ 
        \midrule
        \makecell{\glsentryshort{pimm}~\cite{desnos_pimm_2013}\\(\glsentryshort{pisdf})} & \glsentryshort{wiso} & \glsentryshort{wiso} & $\mathbb{N}$ & \makecell{\glsentryshort{hi} \\ \glsentryshort{mm} \\ \glsentryshort{pa}} & \makecell{\glsentryshort{co}~\cite{desnos_pimm_2013}; \glsentryshort{li}~\cite{desnos_pimm_2013}; \\ \glsentryshort{qsc}~\cite{desnos_pimm_2013}} & N/A\tnote{3} \\ 
        \midrule
        \gls{sad}~\cite{arrestier_delays_2018} & never & never & $\mathbb{N}^*$ & \makecell{\glsentryshort{inidisit} \\ \glsentryshort{mm}} & \makecell{Same static analyzability as\\the underlying \gls{dfmocc}} & N/A\tnote{3} \\
        \bottomrule
      \end{tabular}
      \begin{tablenotes}
        \item[1] for a subclass of \gls{psdf} specification
        \item[2] for a subclass of \gls{pcsdf} specification
        \item[3] those models are meta-model so we don't consider the turing-completeness
        \item \emph{Rate and topology updates acronyms}: \glsentryshort{biso}: \glsentrylong{biso}, \glsentryshort{wiso}: \glsentrylong{wiso}
        \item \emph{Features acronyms}: \glsentryshort{hi}: \glsentrylong{hi}, \glsentryshort{inidisit}: \glsentrylong{inidisit}, \glsentryshort{mm}: \glsentrylong{mm}, \glsentryshort{pa}: \glsentrylong{pa}, \glsentryshort{ph}: \glsentrylong{ph}
        \item \emph{Analyzability acronyms}: \glsentryshort{co}: \glsentrylong{co}, \glsentryshort{li}: \glsentrylong{li}, \glsentryshort{qsc}: \glsentrylong{qsc}
      \end{tablenotes}
    \end{threeparttable}
  \end{table*}

  \subsection{DF MoCCs with Enable and Invoke Capabilities}

  \label{sec:dfmocc-with-enable-and-invoke-capabilities}

  This category includes \glspl{dfmocc} where actors have two capabilities: \emph{enable} and \emph{invoke}. The \emph{enable} capability determines if an actor can execute in a given mode, while the \emph{invoke} capability performs the execution in that mode.

  \subsubsection{EIDF}

  \gls{eidf}~\cite{plishker_heterogeneous_2011} endows actors with two \emph{capabilities} and a set of \emph{modes}. Each mode defines a number of tokens to be produced and consumed. The \emph{enable} capability asserts if an actor can execute in a given mode while the \emph{invoke} capability performs the execution in that mode. Besides the number of produced tokens, the invoke capability returns the set of allowed modes for the subsequent execution of the actor. This set can be empty or contain a single or multiple elements. Especially, as multiple modes can be allowed for the next execution, an actor's behavior may differ depending on the arrival times of the tokens, and \gls{eidf} is non-deterministic.

  \subsubsection{CFDF}

  The invoke capability results in both the output tokens and the set of enabled modes for the subsequent executions. The \gls{cfdf}~\cite{plishker_heterogeneous_2011} behaves the same as \gls{eidf}, except that the invoke capability returns a single mode, which makes \gls{cfdf} deterministic. As the enable and invoke capabilities can be formulated to describe switch/select actors of a \gls{bdf} model, \gls{eidf} and \gls{cfdf} are also Turing-complete.

  \subsubsection{PSM-CFDF}

  \gls{psm-cfdf}~\cite{lin_parameterized_2015} is tailored for \gls{cfdf} when the number of modes grows significantly. Actors have a set of parameters, and a \emph{configuration} is an assignation to those parameters. Modes with related functionalities are clustered together and denoted as \gls{psm}. The active \gls{psm} and the active configuration uniquely determine the mode for the actor executions. There are multiple next \gls{psm} that can be reached. As \gls{psm-cfdf} is not explicitly stated as deterministic, we assume that \gls{psm-cfdf} is non-deterministic.

  \subsubsection{CF-PSDF}

  \gls{cf-psdf}~\cite{wang_parameterized_2013} is a mix between \gls{psdf} and \gls{cfdf}. A \gls{cf-psdf} actor has modes and three graphs: the ctrl graph, the subctrl graph, and the body graph. The ctrl and subctrl graphs have the same role as the init and subinit graphs of \gls{psdf}. The ctrl graph decides the execution mode and transmits the mode information to the ctrl graph of subsequent \gls{cf-psdf} actors. Two distinct actors can control a \gls{cf-psdf} actor. The first sends mode information to the ctrl graph, and the second sends data to the body graph.

  \subsubsection{HCFDF}

  \gls{hcfdf}~\cite{sudusinghe_novel_2013} specifies its actors as \gls{cfdf} actors with a set of nested DFGs. Let $H$ be an \gls{hcfdf} actor. The nested DFGs match a subset of ports of $H$. An execution of $H$ might be an invocation of a subset of the nested graphs, given that the dataflow interface defined by the mode is unchanged.

  \begin{table*}[htbp]
    \centering
    \caption{Features and static analyses of \nameref{sec:dfmocc-with-enable-and-invoke-capabilities}.}
    \label{tab:dfmocc-with-enable-and-invoke-capabilities}
    \begin{threeparttable}[htbp]
      \begin{tabular}{|c|c|c|c|c|c|c|c|}
        \toprule
        \textbf{\glsentryshort{dfmocc}} & \makecell{\textbf{Rate} \\ \textbf{updates}} & \makecell{\textbf{Topology} \\ \textbf{updates}} & \makecell{\textbf{Range} \\ \textbf{rate}} & \textbf{Features} & \textbf{Static analyses} & \makecell{\textbf{Turing} \\ \textbf{complete}} \\
        \midrule
        \gls{eidf}~\cite{plishker_heterogeneous_2011} & \glsentryshort{wiso} & \glsentryshort{wiso} & $\mathbb{N}$ & N/A\tnote{1} & N/A & $\bullet$ \\
        \midrule
        \gls{cfdf}~\cite{plishker_heterogeneous_2011} & \glsentryshort{wiso} & \glsentryshort{wiso} & $\mathbb{N}$ & N/A\tnote{1} & \glsentryshort{det}~\cite{bhattacharyya_handbook_2013} & $\bullet$ \\
        \midrule
        \gls{psm-cfdf}~\cite{lin_parameterized_2015} & \glsentryshort{wiso} & \glsentryshort{wiso} & $\mathbb{N}$ & \glsentryshort{pa} & N/A & $\bullet$ \\
        \midrule
        \gls{cf-psdf}~\cite{wang_parameterized_2013} & \glsentryshort{wiso} & \glsentryshort{wiso} & $\mathbb{N}$ & \makecell{\glsentryshort{hi} \\ \glsentryshort{pa}} & N/A & $\bullet$ \\
        \midrule
        \gls{hcfdf}~\cite{sudusinghe_novel_2013} & \glsentryshort{wiso} & \glsentryshort{wiso} & $\mathbb{N}$ & \glsentryshort{hi} & N/A & $\bullet$ \\
        \bottomrule
      \end{tabular}
      \begin{tablenotes}
        \item \emph{Rate and topology updates acronyms}: \glsentryshort{wiso}: \glsentrylong{wiso}
        \item \emph{Features acronyms}: \glsentryshort{hi}: \glsentrylong{hi}, \glsentryshort{pa}: \glsentrylong{pa}
        \item \emph{Analyzability acronyms}: \glsentryshort{det}: \glsentrylong{det}
      \end{tablenotes}
    \end{threeparttable}
  \end{table*}

  \subsection{Process network-based DF MoCCs}

  \label{sec:process-network-based-dfmocc}

  This category includes dataflow formalism that generalizes the concept of \glspl{dfmocc}. This generalization is referred to as \glspl{dpn}.

  \subsubsection{KPN}

  Actors are sometimes called \gls{dp}, and a network of \gls{dp} is a \gls{dpn}~\cite{lee_dataflow_1995}. When a \gls{dp} executes, it consumes tokens from its input channels and produces tokens on its output channels. A set of \emph{execution rules} indicates when the \gls{dp} is \emph{enabled} to execute. For instance, an actor of a \gls{sdf} specification is enabled when it has enough tokens in its input channels, and an actor of an \gls{rmdf} specification with a frequency constraint is enabled when it has both enough tokens in its input channels and when time is within its execution windows. Dataflow specifications we have studied so far are a particular type of \gls{kpn}~\cite{kahn_semantics_1974}. A \gls{kpn} is a collection of concurrent processes that communicate with channels. In contrast with \gls{dp} processes, \gls{kpn} processes cannot test for the presence or absence of tokens in a given input channel. \gls{kpn} processes are also \emph{continuous} rather than discrete: there is no well-defined notion of the quantity of computation. Thus, input tokens are processed as soon as they become available, and output tokens are processed as soon as they are produced. A \gls{kpn} is a deterministic system~\cite{kahn_semantics_1974}.

  \subsubsection{RPN}

  \glspl{rpn}~\cite{geilen_reactive_2004} is an extension of \glspl{kpn} where the set of active processes and channels may change at \rt on receiving events, which introduced non-determinism compared to \glspl{kpn}. An \gls{rpn} presents a static interface to the outside world that receives events and tokens.~Events introduce non-functional determinism in the model.

  \begin{table*}[htbp]
    \centering
    \caption{Features and static analyses of \nameref{sec:process-network-based-dfmocc}.}
    \label{tab:process-network-based-dfmocc}
    \begin{threeparttable}[htbp]
      \begin{tabular}{|c|c|c|c|c|c|c|}
        \toprule
        \textbf{\gls{dfmocc}} & \makecell{\textbf{Rate} \\ \textbf{updates}} & \makecell{\textbf{Topology} \\ \textbf{updates}} & \makecell{\textbf{Range} \\ \textbf{rate}} & \textbf{Features} & \textbf{Static Analyses} & \makecell{\textbf{Turing} \\ \textbf{Complete}} \\
        \midrule
        \glsentryshort{kpn}~\cite{kahn_semantics_1974} & \glsentryshort{wiso} & \glsentryshort{wiso} & $\Omega$\tnote{1} & N/A & \glsentryshort{det}~\cite{kahn_semantics_1974} & $\bullet$ \\
        \midrule
        \glsentryshort{rpn}~\cite{geilen_reactive_2004} & \glsentryshort{wiso} & \glsentryshort{wiso} & $\Omega$\tnote{1} & \glsentryshort{hi} & N/A & $\bullet$ \\
        \bottomrule
      \end{tabular}
      \begin{tablenotes}
        \item[1] any type of objects (pointers, integers, functions, etc); the choice of \enquote{omega} is inspired from the probability theory where it represents the set of all possible outcomes
        \item \emph{Rate and topology updates acronyms}: \glsentryshort{wiso}: \glsentrylong{wiso}
        \item \emph{Features acronyms}: \glsentryshort{hi}: \glsentrylong{hi}
        \item \emph{Analyzability acronyms}: \glsentryshort{det}: \glsentrylong{det}
      \end{tablenotes}
    \end{threeparttable}
  \end{table*}

  \section{Expressiveness and Analyzability Hierarchy}

  \label{sec:hierarchy}

  \subsection{Protocol for Creating Expressiveness and Analyzability Hierarchy}

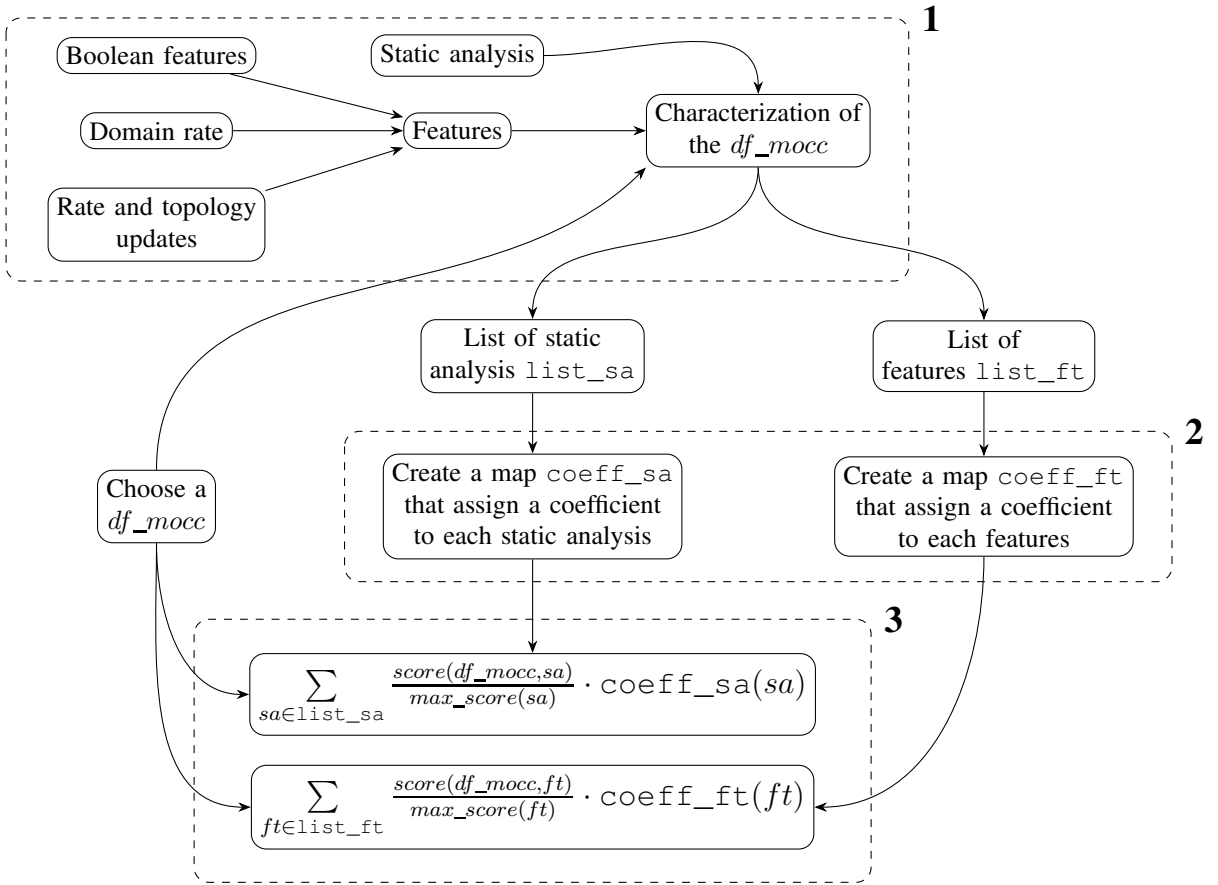
\begin{figure*}[htbp]
    \centering
    \begin{tikzpicture}
      \node[actor] (boolean) at (0,0) {Boolean features};
      \node[actor] (domain-rate) at (0,-1) {Domain rate};
      \node[actor] (rate-and-topology) at (0,-2.25) {Rate and topology \\ updates};
      \node[actor] (sa) at (4,0) {Static analysis};
      \node[actor] (features) at (4,-1) {Features};
      \node[actor] (characterization) at (8,-1) {Characterization of \\ the $df\_mocc$};
      \node[actor] (list-sa) at (5,-4) {List of static \\ analysis \texttt{list\_sa}};
      \node[actor] (list-ft) at (11,-4) {List of \\ features \texttt{list\_ft}};
      \node[actor] (node1) at (5,-6) {Create a map \texttt{coeff\_sa} \\ that assign a coefficient \\ to each static analysis};
      \node[actor] (node2) at (11,-6) {Create a map \texttt{coeff\_ft} \\ that assign a coefficient \\ to each features};
      \node[actor] (score-sa) at (5,-8.5) {\large$\sum\limits_{sa \in \texttt{list\_sa}} \frac{score(df\_mocc, sa)}{max\_score(sa)} \cdot \texttt{coeff\_sa}(sa)$};
      \node[actor] (score-ft) at (5,-10) {\large$\sum\limits_{ft \in \texttt{list\_ft}} \frac{score(df\_mocc, ft)}{max\_score(ft)} \cdot \texttt{coeff\_ft}(ft)$};
      \node[actor] (dfmocc) at (0,-6) {Choose a \\ $df\_mocc$};
      \node at (10.3,0.5) {\Large\textbf{1}};
      \node at (13.8,-5) {\Large\textbf{2}};
      \node at (9.8,-7.5) {\Large\textbf{3}};
      \draw[black,dashed,rounded corners=2mm] (-2,0.5) rectangle (10,-3);
      \draw[black,dashed,rounded corners=2mm] (2.5,-5) rectangle (13.5,-7);
      \draw[black,dashed,rounded corners=2mm] (0.5,-7.5) rectangle (9.5,-11);
      \draw[arc] (boolean) to (features);
      \draw[arc] (domain-rate) to (features);
      \draw[arc] (rate-and-topology) to (features);
      \draw[arc] (features) to (characterization);
      \draw[arc] (characterization) to[out=270,in=90] (list-sa);
      \draw[arc] (characterization) to[out=270,in=90] (list-ft);
      \draw[arc] (list-sa) to (node1);
      \draw[arc] (list-ft) to (node2);
      \draw[arc] (node1) to (score-sa);
      \draw[arc] (node2) to[out=270,in=0] (score-ft);
      \draw[arc] (dfmocc) to[out=270,in=180] (score-sa);
      \draw[arc] (dfmocc) to[out=270,in=180] (score-ft);
      \draw[arc] (dfmocc) to[out=90,in=225] (characterization.south west);
      \draw[arc] (sa) to[out=0,in=90] (characterization);
    \end{tikzpicture}
    \caption{Visualization of the protocol to compute the expressiveness and analyzability hierarchy.}
    \label{fig:protocol-visualization}
  \end{figure*}

  The classification in the previous section can be used to quantitatively compare the expressiveness and analyzability of \glspl{dfmocc}. We call this the \emph{expressiveness hierarchy} and the \emph{analyzability hierarchy}. The protocol to create those hierarchies is the following, and they are also illustrated in \cref{fig:protocol-visualization}:

\begin{enumerate}
  \item Characterize each \gls{dfmocc} according to the features described in \cref{sec:features} and static analyzability described in \cref{sec:analyzability}; this has been done in this paper (cf. \cref{tab:sdf-and-related-dfmoccs} to \cref{tab:process-network-based-dfmocc}).
  \item Assign a coefficient to each feature and static analysis according to the designer's need. Those coefficients aim to increase or reduce the importance of features or static analysis. For instance, coefficient 1 can be used for needed features and static analysis, and 0 for unneeded features.
  \item Compute the expressiveness and analyzability score for each \gls{dfmocc} by summing the normalized features' score with the correct weighting, then sort \gls{dfmocc} regarding their expressiveness and analyzability score.
  \begin{itemize}
    \item The score of boolean-valued features is 1 if the feature is present and 0 otherwise.
    \item The score of range rate is 0 if it is $\{1\}$, 1 if $\mathbb{N}^*$, 2 if $\mathbb{N}$, 3 if $\mathbb{Q}^*$, and 4 if $\Omega$. The incrementation by 1 represents an increase in expressiveness.
    \item The rate and topology updates score is computed in \cref{tab:rate-topology-updates-score}. This score is the sum of the score of the rate updates and the topology updates. Each of them is evaluated as follows. If it is evaluated as a single type \eg \gls{wiso} for rate updates of \polygraph, the score is the intersection of the type with itself on \cref{tab:rate-topology-updates-score}, that is 6. Thus, the score for rate updates for \polygraph is 6. If it is evaluated as two different types \eg \gls{wiso} and \gls{wiro} for \gls{rmdf}, the score is the intersection of the two types, that is 7. Thus, the score for rate updates for \gls{rmdf} is 7.
  \end{itemize}
\end{enumerate}

\begin{table}[htbp]
  \centering
  \caption[Evaluation of rates and topology updates]{Evaluation of rates and topology updates.}
  \label{tab:rate-topology-updates-score}
  \begin{tabular}{|c|c|c|c|c|c|}
    \toprule
    & never & \glsentryshort{biso} & \glsentryshort{biro} & \glsentryshort{wiso} & \glsentryshort{wiro} \\
    \midrule
    never & 0 & 1 & 2 & 3 & 4 \\
    \midrule
    \glsentryshort{biso} & 1 & 2 & 3 & 4 & 5 \\
    \midrule
    \glsentryshort{biro} & 2 & 3 & 4 & 5 & 6 \\
    \midrule
    \glsentryshort{wiso} & 3 & 4 & 5 & 6 & 7 \\
    \midrule
    \glsentryshort{wiro} & 4 & 5 & 6 & 7 & 8 \\
    \bottomrule
  \end{tabular}
\end{table}

  \subsection{Application of the Protocol}

  Let us assume that we are interested in specifying and analyzing \glspl{cps} with relaxed real-time constraints and mode-dependent execution. \glspl{cps} with relaxed real-time constraints have real-time constraints on only a subset of their processes, and \glspl{cps} with a mode-dependent execution have conditional execution branches. Thus, we are interested in the \emph{Rate and topology updates} features to specify mode-dependent executions of \glspl{cps} as much as possible. Regarding specifying real-time constraints, we are interested in the features \emph{\glsentrylong{del}}, \emph{\glsentrylong{et}}, and \emph{\glsentrylong{freq}}. In addition, the features \emph{Domain rate}, \emph{\glsentrylong{ph}} and \emph{\glsentrylong{it}} are also interesting in easing the specification of real-time components with different frequencies.
  
  Regarding the static analysis, let us assume that we want the \gls{dfmocc} to be \emph{deterministic} and to provide a \emph{consistency} and \emph{liveness} analysis. The analysis of \emph{execution windows} would also be of high interest.
  
\begin{figure*}
    \begin{tikzpicture}
      \begin{axis}[xmin=-1,
        xmax=7.5,
        ymin=-0.5,
        ymax=7.5,
        xtick={0,1,2,...,7},
        width=\textwidth,
        height=0.5\textwidth,
        ytick={0,1,...,5},
        axis lines = left,
        ylabel=Analyzability score,
        xlabel=Expressiveness score]
          \addplot[color=black, mark=Mercedes star, only marks] coordinates {(0, 3) (0.25,3) (0.5,3) (1,3) (1.25,3) (1.5,3) (1.5,2) (1.75, 2) (2,2) (2.25,3) (3,2) (3,3) (3,1) (3.5, 0) (3.5, 2) (3.75,2) (4,3) (5.25, 3) (6.25, 4) (6.5,4)};

          \node[left] at (axis cs: 0, 3) {HSDF};
          \node[align=center,draw] (legend) at (axis cs: 0.25, 5.5) {CV-SDF WSDF \\ IBSDF SPBDF \\ MDSDF SSDF \\ VSDF SDF CG};
          \node[below] at (axis cs: 0.5, 3) {HDF};
          \node[align=center,draw] (legend2) at (axis cs: 2, 6) {HSDF$^a$ \\ SPDF};
          \node[align=center,draw] (legend3) at (axis cs: 0.25, 1) {ILDF \\ tSDF RDF};
          \node[align=center,draw] (legend4) at (axis cs: 2, 0.5) {FSM-SADF \\ FSM-PSADF};
          \node[align=center,draw] (legend5) at (axis cs: 4, 5) {SADF \\ eSADF};
          \node[align=center,draw] (legend6) at (axis cs: 4.5, 0.5) {VRDF \\ VPDF};
          \node[align=center,draw] (legend7) at (axis cs: 6, 6) {Dyn. \polygraph \\ TPDF};
          \node[align=center,draw] (legend8) at (axis cs: 5.75, 1.5) {tCSDF \\CSDF$^a$};
          \node[above] at (axis cs: 1.5, 3) {PCG};
          \node[below] at (axis cs: 1.5, 2) {BDDF};
          \node[above] at (axis cs: 2, 2) {BPDF};
          \node[above] at (axis cs: 2.25, 3) {ppSDF};
          \node[above] at (axis cs: 3, 3) {CSDF};
          \node[above] at (axis cs: 3.5, 0) {xSADF};
          \node[above] at (axis cs: 3.5, 2) {CDDF};
          \node[below] at (axis cs: 3.75, 2) {FRDF};
          \node[below] at (axis cs: 6.25, 4) {\polygraph};
          \node[right] at (axis cs: 6.5, 4) {RMDF};
  
          \draw (legend) -- (axis cs: 0.25, 3);
          \draw (legend2) to[out=-90,in=90] (axis cs: 1, 3);
          \draw (legend3) to[out=45,in=-150] (axis cs: 1.25, 3);
          \draw (legend4.north) to[out=45,in=-30] (axis cs: 1.75, 2);
          \draw (legend5) to[out=-90,in=90] (axis cs: 3, 2);
          \draw (legend6.north west) to[out=180,in=-30] (axis cs: 3, 1);
          \draw (legend7) to[out=180,in=90] (axis cs: 5.25, 3);
          \draw (legend8) to[out=135,in=330] (axis cs: 4, 3);
      \end{axis}
    \end{tikzpicture}
    \caption{Expressiveness and analyzability score of \glspl{dfmocc} which are non-Turing complete and non-meta-model ones. The coefficient used to compute scores is 1 for the useful features and 0 for the unneeded ones.}
    \label{fig:expressiveness-analyzability-score}
  \end{figure*}
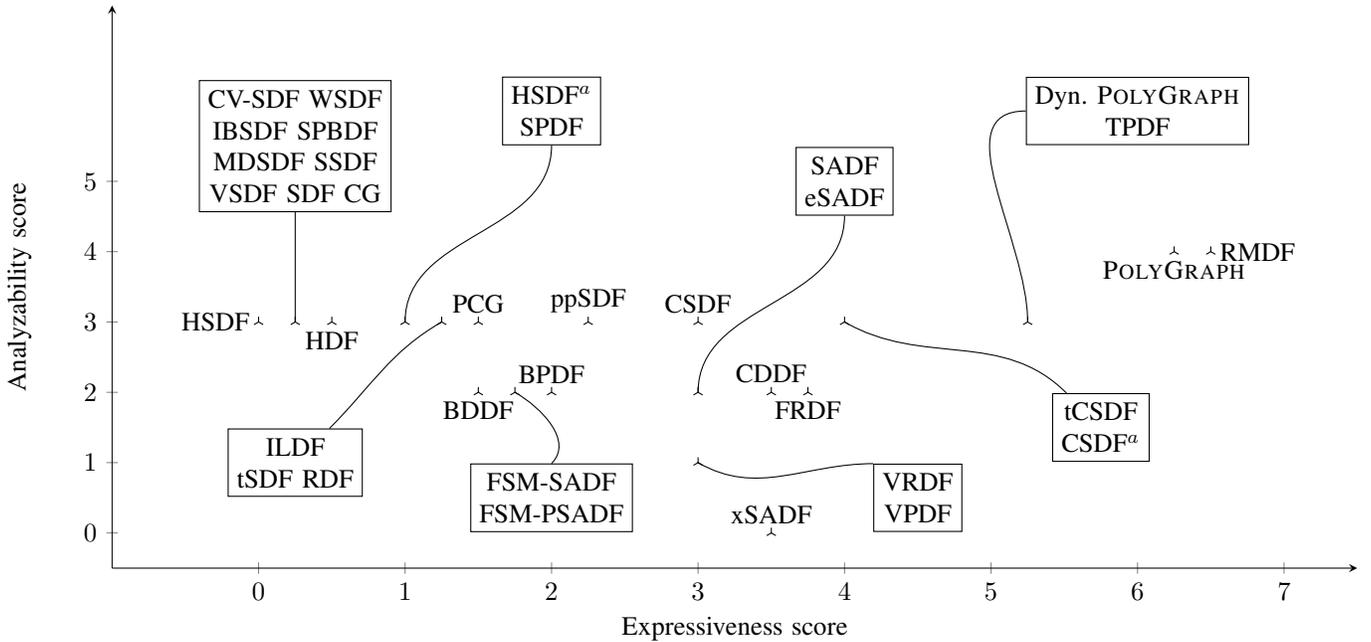

  The \cref{fig:expressiveness-analyzability-score} shows the expressiveness and analyzability score where we assign coefficient 1 for the features and static analyses of interest (as defined in previous paragraphs) of interest and coefficient 0 for the other. \glspl{dfmocc} shown in \cref{fig:expressiveness-analyzability-score} are the non-Turing complete and non-meta-model \gls{dfmocc} studied in this paper. We exclude Turing-complete and meta-model \gls{dfmocc} from this analysis as the static analysis of the former is limited, and the latter is not intended to be used on its own. We can see in \cref{fig:expressiveness-analyzability-score} that \polygraph, Dynamic \polygraph, \gls{tpdf} and \gls{rmdf} are the most suitable \gls{dfmocc} to specify and analyze \glspl{cps} with relaxed real-time constraints and mode-dependent execution

  \subsection{Extending the Classification}

  New features, static analyses, and \glspl{dfmocc} can be easily integrated into the classification system. A feature is introduced by defining how it is evaluated and evaluating each \gls{dfmocc} against this feature. A new static analysis is incorporated by specifying what it evaluates and evaluating each \gls{dfmocc} against this static analysis. A \gls{dfmocc} is added by characterizing it with the existing features and static analyses. Dependencies, whether among features or static analyses, can be seamlessly integrated.

  \subsection{A Visualization Tool}

  To enhance the accessibility of our classifiation system, we have developed an open-source visualization tool\footnote{https://github.com/groumage/DFMoCCs-survey}. This tool facilitates the comparative evaluation of \glspl{dfmocc} and enable system designers and researchers to filter \glspl{dfmocc} based on required features and static analyses. This contribution aims to support engineers and researchers in navigating the broad ecosystem of \glspl{dfmocc}.

  \section{Conclusion}

  \label{sec:conclusion}

  We have presented a survey and a comparison framework for \glspl{dfmocc} found in the scientific literature. Our work includes a comprehensive list of features and static analyses designed to characterize the expressiveness and the analyzability of each \gls{dfmocc}. Building on this characterization, we proposed a protocol to assign expressiveness and analyzability scores to each \gls{dfmocc} according to system designer needs. The framework we proposed is easily extensible, allowing for the incorporation of new features, static analyses, and dataflow models. Our classification quantitatively supports a widely accepted assertion in the \gls{dfmocc} community: there is a trade-off between expressiveness and analyzability in \glspl{dfmocc}. The framework we proposed can serve both as a comparative tool and as a decision-making aid. We hope that the \gls{dfmocc} community will participate in this effort by providing feedback and by extending the classification system through the GitHub repository of the paper\footnote{https://github.com/groumage/DFMoCCs-survey}.

  \bibliographystyle{IEEEtran}

\end{document}